\documentclass[11pt,a4paper]{article}
\pdfoutput=1
\usepackage{jheppub}

\usepackage[utf8]{inputenc}
\usepackage{graphicx}
\usepackage{amsmath,amsfonts,amssymb}
\usepackage{epsfig,color}

\usepackage{slashed,setspace,float}
\usepackage{bm,latexsym}
\usepackage[outdir=./]{epstopdf}
\usepackage[T1]{fontenc}

\usepackage{dcolumn}

\usepackage[subfigure]{graphfig}

\usepackage{amsxtra}
\usepackage{hyperref}
\usepackage{placeins}

\usepackage{amssymb,amsthm}
\usepackage{wasysym}
\usepackage{hyperref}
\usepackage{braket}

\usepackage[shortlabels]{enumitem}
\usepackage{gensymb,verbatim}
\usepackage{cancel}
\usepackage{multirow}
\usepackage{xcolor}
\usepackage{mathrsfs}

\usepackage{physics}

\newcommand{\euv}{\varepsilon_{\rm UV}}

\newcommand{\nn}{\nonumber \\ }

\allowdisplaybreaks[4]

\hypersetup{colorlinks, linkcolor = [rgb]{0,0.0,1.0}, citecolor = [rgb]{0,0.0,1.0}, urlcolor = [rgb]{0,0.0,1.0}}

\begin{document}

\title{Polarized Gluon   Pseudodistributions  at Short Distances}

\author{Ian Balitsky, Wayne Morris and  Anatoly  Radyushkin}
\affiliation{Old Dominion University, Norfolk, \\4600 Elkhorn Ave., Norfolk, VA 23529, USA
            }
\affiliation{Thomas Jefferson National Accelerator Facility,
              \\ 12000 Jefferson Ave., Newport News, VA 23606, USA
}

    \emailAdd{balitsky@jlab.org}
 \emailAdd{wmorr001@odu.edu}
                \emailAdd{radyush@jlab.org}
                        
\abstract{
We  formulate the basic points 
of the pseudo-PDF approach to the lattice  calculation of polarized  gluon PDFs.  
We  present the   results  of our calculations of the one-loop 
corrections for the bilocal  $G_{\mu \alpha}(z)  \widetilde G_{\lambda \beta}(0)$ correlator of gluonic fields. 
Expressions are given  for  a general  situation when all four indices are arbitrary,
and also for specific  combinations of indices corresponding to     three matrix elements
that contain  the   twist-2 invariant amplitude related to the polarized PDF. 
We   study  the evolution properties  of  these matrix elements,
and derive matching relations between Euclidean and light-cone Ioffe-time distributions.
These relations 
 are necessary for extraction of the polarized gluon distributions from the lattice data.

}
                        
\maketitle

\newpage

\section{
 Introduction}

Lattice calculations  devoted to  extraction   of  the parton distribution functions (PDFs) have attracted recently  a  considerable interest
(see Refs.  \cite{Constantinou:2020hdm,Cichy:2018mum}  for   reviews and references).
Starting with the paper  \cite{Ji:2013dva} by   X. Ji,  modern efforts  aim at directly getting  PDFs $f(x)$  as functions 
of the momentum fraction variable $x$  rather than just calculating 
their $x^N$ moments.
The key element of these efforts   is the analysis of equal-time bilocal operators 
 that define various parton functions, in particular,   PDFs, distribution amplitudes (DAs), generalized  parton distributions (GPDs), and 
 transverse momentum dependent distributions (TMDs). 
The   major object 
 of Ji's approach in the case of ordinary PDFs,   
 are     quasi-PDFs $Q(y,p_3)$  \cite{Ji:2013dva,Ji:2014gla}. 
To get the 
PDFs  from them, one should take    the  large-momentum $p_3 \to \infty$  limit 
of  $Q(y,p_3)$. 

There are alternative  methods based on the  coordinate-space formulation, such as  
 the ``good lattice cross sections'' approach
 \cite{Ma:2014jla,Ma:2017pxb} and the pseudo-PDF approach \cite{Radyushkin:2017cyf,Radyushkin:2017sfi,Orginos:2017kos},
 in which the equal-time correlators  $M(z_3,p_3)$ are considered as functions of the 
 Ioffe-time    \cite{Braun:2007wv,Bali:2017gfr,Bali:2018spj} $\nu=z_3p_3$  and the probing scale parameter $z_3^2$.
In these latter cases, the parton distributions  are extracted by   taking  the 
short-distance $z_3^2 \to 0$ limit at fixed $\nu$.

To  convert the data measured on a     Euclidean  lattice into
the    PDFs defined on the light cone,  it should be  taken into account that the limits 
$p_3\to \infty$ and $z_3\to 0$ are  singular.
 To perform the conversion in such a situation, 
one needs to derive and use   {\it matching relations}.

 In the quasi-PDF approach, the matching relations   were derived 
for quark    \mbox{\cite{Ji:2013dva,Xiong:2013bka,Ji:2015jwa,Izubuchi:2018srq}} and gluon 
 PDFs \cite{Wang:2017eel,Wang:2017qyg,Wang:2019tgg}, and also  for GPDs
\cite{Ji:2015qla,Xiong:2015nua,Liu:2019urm}  and the
 pion DA   \cite{Ji:2015qla}.

The  matching relations 
 for the bilocal operators in the coordinate representation 
were  originally derived in   applications to quark nonsinglet
PDFs   \cite{Ji:2017rah,Radyushkin:2017lvu,Radyushkin:2018cvn,Zhang:2018ggy,Izubuchi:2018srq}.
The pseudo-PDF procedure for   lattice extraction  of nonforward 
parton functions, such as  nonsinglet GPDs and the pion DA
were described 
in  Ref.   \cite{Radyushkin:2019owq},
where the  necessary  matching conditions  
were also obtained. 

  The  pseudo-PDF  approach to the extraction of unpolarized gluon PDFs was formulated 
  in our paper \cite{Balitsky:2019krf} (see also Ref. \cite{Balitsky:2021bds}).
  The results of one-loop calculations for the gluon bilocal operators were presented there,
  and, in  a more detailed form in Ref. \cite{Balitsky:2021qsr}. 
The matching conditions  following from these  results have been used 
in lattice extractions of the unpolarized gluon PDFs 
in Refs.  \cite{Fan:2020cpa,Fan:2021bcr}  and  \cite{HadStruc:2021wmh}. 
One-loop corrections to the  matrix element of the twist-4 ``gluon condensate'' operator
$G^{\mu \nu}(0) G_{\mu \nu} (z)$ have been recently obtained in the momentum-representation 
calculation of Ref. 
\cite{Radyushkin:2021fel}.

In the present work,  we  describe the basics of the pseudo-PDF approach to lattice extraction of the polarized gluon PDFs.
The paper is organized as follows. 
In 
 \mbox{Section  2},   we  investigate kinematic   structure  of the polarized matrix  elements of the gluonic bilocal operators
 built from the gluon stress-tensor and its dual. 
 In particular, we identify the matrix elements 
 that contain information about the twist-2 polarized gluon PDF.
   In \mbox{Section 3,}  we  present the results for  one-loop corrections to the bilocal operator,  and  
   discuss  their  ultraviolet  and short-distance behavior. 
 The matching relations necessary for the lattice extraction of the polarized gluon PDFs 
 are derived in  Section 4.  
  The summary of the paper is given in section 5.


\section{Matrix elements}

\subsection{Definitions}

To extract polarized gluon distributions of a  nucleon, we  consider 
  matrix elements of bilocal operators {$G_{\mu \alpha}(z) \widetilde G_{ \lambda \beta } (0)$ composed}  of  two  gluon fields,    with
the dual field defined by   \mbox{$ \widetilde G_{ \lambda \beta }  = \frac12 \epsilon_{\lambda \beta \rho \gamma} G^{\rho \gamma}$. } 
 The matrix elements  are   specified  by 
    \begin{align}
\widetilde { m}_{\mu \alpha;  \lambda \beta }  (z,p) \equiv \langle  p,s |  \,
 G_{\mu \alpha} (z)  \,  { \tilde E} (z,0; A) \widetilde G_{ \lambda \beta } (0) | p,s \rangle \  , 
\end{align}
 where  $
{ \tilde E}(z,0; A)$ is  the  usual  $0\to z$ straight-line gauge link 
 in the gluon  (adjoint) 
 representation
 \begin{align}
{ \tilde E}(z,0; A) \equiv P \exp \left [ ig \,  z_\sigma\, \int_0^1d t \,  \tilde   A^\sigma (t z) 
 \right ]    \  . 
 \label{straightE}
\end{align}
The standard  definition of the polarized gluon PDFs  \cite{Manohar:1990jx} uses  
the contracted amplitude  $g^{\alpha \lambda} { m}_{\mu \alpha;  \lambda \beta }$, but   we will keep    all
 four indices  $\mu, \alpha, \lambda, \beta$  non-contracted. 
The  part that  depends  on the nucleon spin is determined by the $z$-odd combination,  
which  vanishes for  the unpolarized case and  is    linear in the spin-vector  $s$. Thus,  we  start 
with the amplitude 
 \begin{align}
 \widetilde{ M}_{\mu \alpha;   \lambda  \beta} (z,p) \equiv  \widetilde{ m}_{\mu \alpha;  \lambda \beta }  (z,p) - 
 \widetilde{ m}_{\mu \alpha;  \lambda \beta }  (-z,p) \ .
\end{align}
 To simplify further   formulas, we  
normalize  $s_\mu$ by 
\mbox{$s^2=-m^2$}, where $m$ is the nucleon mass. This means that our polarization
vector $s_\mu$ is related by $s_\mu = m S_\mu$ to the usual 
polarization vector $S_\mu$ which is normalized by $S^2=-1$.

\subsection{Invariant amplitudes}

  The tensor structures for   the   decomposition of 
  $ \widetilde{ M} _{\mu \alpha;   \lambda  \beta}  (z,p) $ over   invariant amplitudes    
 may be built from two   available 4-vectors $p_\alpha$, $z_\alpha$, 
 one pseudo-vector $s_\alpha$   and the 
 metric tensor $g_{\alpha \beta}$.
These structures must be  anti-symmetric with respect to  interchange of 
both $\{\mu \leftrightarrow \alpha \}$ and $\{\lambda  \leftrightarrow \beta \}$.  
    
 Let us list  first the structures in which $s$ carries one of the ${\mu \alpha;   \lambda  \beta}$ indices.
Such structures,  before the  anti-symmetrization,   may have two possible forms:  $s_\alpha A_\beta  g_{\mu\lambda} $ and 
 $ s_\alpha A_\beta B_{\mu} C_{\lambda}$, where $A, B, C$ correspond to  $p$ or $z$. 
 Incorporating the antisymmetry   of $G_{\rho \sigma}$ with  respect to its indices, we have 
      \begin{align}
 \widetilde{ M} _{\mu \alpha;   \lambda  \beta}^{(1)}  (z,p)= &
 \left( g_{\mu\lambda} s_\alpha p_\beta - g_{\mu\beta} s_\alpha p_\lambda - g_{\alpha\lambda} s_\mu p_\beta + g_{\alpha\beta} s_\mu p_\lambda \right) \widetilde{\mathcal{M}}_{sp}    \nonumber \\
+ & \left( g_{\mu\lambda} p_\alpha s_\beta - g_{\mu\beta} p_\alpha s_\lambda - g_{\alpha\lambda} p_\mu s_\beta + g_{\alpha\beta} p_\mu s_\lambda \right) \widetilde{\mathcal{M}}_{ps}    \nonumber \\ +& \left( g_{\mu\lambda} s_\alpha z_\beta - g_{\mu\beta} s_\alpha z_\lambda - g_{\alpha\lambda} s_\mu z_\beta + g_{\alpha\beta} s_\mu z_\lambda \right) \widetilde{\mathcal{M}}_{sz}    \nonumber \\
+ & \left( g_{\mu\lambda} z_\alpha s_\beta - g_{\mu\beta} z_\alpha s_\lambda - g_{\alpha\lambda} z_\mu s_\beta + g_{\alpha\beta} z_\mu s_\lambda \right) \widetilde{\mathcal{M}}_{zs}    \nonumber \\ 
+&   (p_\mu s_\alpha -p_\alpha s_\mu ) (p_\lambda z_\beta -p_\beta z_\lambda)   \widetilde{\mathcal{M}}_{pspz}   
+   ( p_\mu z_\alpha-p_\alpha z_\mu )( p_\lambda s_\beta - p_\beta s_\lambda) 
 \widetilde{\mathcal{M}}_{pzps}    \nonumber \\
+&   ( s_\mu z_\alpha - s_\alpha z_\mu) (p_\lambda z_\beta -  p_\beta z_\lambda)
\widetilde{\mathcal{M}}_{szpz}  
+( p_\mu z_\alpha  - p_\alpha z_\mu )(s_\lambda z_\beta -s_\beta z_\lambda) \widetilde{\mathcal{M}}_{pzsz}  \ ,
\label{M1}
\end{align}
where the invariant amplitudes $\widetilde{\mathcal{M}}$ are functions of 
the invariant interval $z^2$ and the Ioffe time \cite{Braun:1994jq}
$(pz)\equiv - \nu$   (the minus sign here
is introduced to have \mbox{$\nu= p_3 z_3$}  when \mbox{$z=\{0,0,0,z_3 \}$).}

There are  also  structures  containing $s$  through  the $(sz)$  product 
accompanied by  all the tensor combinations
of $p,z$ and metric tensor that have been used in Ref. \cite{Balitsky:2019krf}  for  the unpolarized case.
These combinations, before   the anti-symmetrization,   may have three possible forms:  
$A_\alpha B_\beta  g_{\mu\lambda} $, 
 $ A_\alpha B_\beta C_{\mu} D_{\lambda}$  and $g_{\alpha\beta} g_{\mu\lambda} $, 
 where $A, B, C,D$ correspond to one of  $p$ or $z$.  Thus, we have 
   \begin{align}
  \widetilde{ M} _{\mu \alpha;   \lambda  \beta}^{(2)}  (z,p)=& 
 (sz)\left( g_{\mu\lambda} p_\alpha p_\beta - g_{\mu\beta} p_\alpha p_\lambda - g_{\alpha\lambda} p_\mu p_\beta + g_{\alpha\beta} p_\mu p_\lambda \right) \widetilde{\mathcal{M}}_{pp}  \nonumber \\
+& (sz)\left( g_{\mu\lambda} z_\alpha z_\beta - g_{\mu\beta} z_\alpha z_\lambda - g_{\alpha\lambda} z_\mu z_\beta + g_{\alpha\beta} z_\mu z_\lambda \right) \widetilde{\mathcal{M}}_{zz}     \nonumber \\
+ &(sz) \left( g_{\mu\lambda} z_\alpha p_\beta - g_{\mu\beta} z_\alpha p_\lambda - g_{\alpha\lambda} z_\mu p_\beta + g_{\alpha\beta} z_\mu p_\lambda \right) \widetilde{\mathcal{M}}_{zp}    \nonumber \\
+ &(sz) \left( g_{\mu\lambda} p_\alpha z_\beta - g_{\mu\beta} p_\alpha z_\lambda - g_{\alpha\lambda} p_\mu z_\beta + g_{\alpha\beta} p_\mu z_\lambda \right) \widetilde{\mathcal{M}}_{pz}     \nonumber \\
+& (sz)  \left( p_\mu z_\alpha  - p_\alpha z_\mu\right) \left(  p_\lambda z_\beta -  p_\beta z_\lambda\right) \widetilde{\mathcal{M}}_{ppzz} 
\nn  + & (sz)  \left(g_{\mu\lambda} g_{\alpha\beta} -g_{\mu\beta} g_{\alpha\lambda} \right)\widetilde{\mathcal{M}}_{gg}    \ .
\label{M2}
\end{align}
 
 One may propose to check if we may also use   the Levi-Civita tensor like $\epsilon_{\gamma \delta \rho \sigma}$  
for building possible tensor structures. Here we note that our matrix element  $\widetilde{ M} _{\mu \alpha;   \lambda  \beta}$
is a pseudo-tensor.  Furthermore, it should be linear in the nucleon polarizations vector $s_\gamma$,
which is  a pseudo-vector.  Hence, the  Levi-Civita pseudo-tensor $\epsilon_{\star \star\star\star}$  
should appear twice in a particular tensor structure involving $s_\gamma$. However, the product of two Levi-Civita tensors $\epsilon_{\star\star\star\star}\epsilon_{\star\star\star\star} $
may be always written in terms of (sums of products of) metric tensors $g_{\star\star}$\ . 
Thus, the combinations  listed in Eqs. (\ref{M1}) and (\ref{M2}) exhaust all the possibilities for {tensor structures}
compliant with the  Lorentz covariance and antisymmetry 
  of  $G_{\rho \sigma}$ with  respect to its indices.
  
In fact, our  operator  has the structure $\epsilon_{\lambda \beta \rho \gamma} 
G^{\lambda \beta}(z)  G^{\rho \gamma}(0)$, where $ G(z)$ and $G(0)$ is the same field.
As we will see in Sect. (\ref{2.5}), this imposes two relations (\ref{SR}),  (\ref{SR2})  between some  invariant 
amplitudes  $\widetilde{\mathcal{M}}$ parametrizing   $\widetilde{ M} _{\mu \alpha;   \lambda  \beta}^{(2)} $
and invariant 
amplitudes entering into   $\widetilde{ M} _{\mu \alpha;   \lambda  \beta}^{(1)} $. 
One may also incorporate the symmetry properties of $ \widetilde{ M} _{\mu \alpha;   \lambda  \beta}  (z,p) $
with respect to $z$. Namely,
since $ \widetilde{ M} _{\mu \alpha;   \lambda  \beta}  (z,p) $ is odd in $z$, the invariant amplitudes
$\widetilde{\mathcal{M}}_{sp}, \widetilde{\mathcal{M}}_{ps}, \widetilde{\mathcal{M}}_{pzsz}, \widetilde{\mathcal{M}}_{szpz}$, $\widetilde{\mathcal{M}}_{zp},
   \widetilde{\mathcal{M}}_{pz}$ are odd functions of $\nu$, while  the remaining ones are even 
functions of $\nu$.

Such a   decomposition  of $ \widetilde{ M} _{\mu \alpha;   \lambda  \beta}  (z,p) $ is quite general.
But it may be also constructed, in particular,  from a  formal Taylor 
expansion of  $ G_{\mu \alpha} (z)  \,  { \tilde E} (z,0; A) \widetilde G_{ \lambda \beta } (0)$
over local operators, followed by  taking their matrix elements and then recombining back 
the terms with the same tensor structure.
The implicit   assumption of this procedure  is that such a Taylor expansion exists. 

In  QCD,  
$ \widetilde{ M}_{\mu \alpha;   \lambda  \beta} (z,p)$ 
has  singularities 
 on the light cone $z^2=0$ due to  perturbative logarithms
$\ln (-z^2)$ generated by gluonic corrections.
Thus, we will assume that   the invariant amplitudes 
$\widetilde{\mathcal{M}} (\nu, z^2)$  are finite for $z^2=0$   at  the tree level,  
and will explicitly calculate the perturbative one-loop corrections that
produce the  $\ln (-z^2)$ terms.

\subsection{Relation to PDF}

The usual light-cone polarized gluon distribution $\Delta g (x)$
is obtained  \cite{Manohar:1990jx} from the matrix element
 $ g^{\alpha \beta} \widetilde{ M}_{+ \alpha;  \beta  +}  (z,p) $,  with $z$ taken in the light-cone ``minus'' direction,
 $z=z_-$.  In terms of the parametrization written above, we  have 
  \begin{align}
 g^{\alpha \beta} &\widetilde{ M}_{+ \alpha;  \beta +}  (z_-,p)=  -2  p_+ s_+ 
 \left [  \widetilde{\mathcal{M}}_{ps}^{(+)}  (\nu ,0)  +  p_+ z_- \widetilde{\mathcal{M}}_{pp} (\nu ,0) \right ] \ ,
 \label{lcpp}
\end{align}
where    ${ \widetilde{\mathcal{M}}}_{ps}^{(+)}  \equiv{ \widetilde{\mathcal{M}}}_{ps} +\widetilde{\mathcal{M}}_{sp}$.
Thus,   the  PDF  is  determined by the structure 
  \begin{align}
   { \widetilde{\mathcal{M}}}_{ps}^{(+)}-
 \nu  \widetilde{\mathcal{M}}_{pp} \equiv -i  {\cal I}_p (\nu) \ .
 \label{Ipnu}
 \end{align}
More specifically, 
  \begin{align}
{\cal I}_p (\nu)  =\frac{i}{2}
 \int_{-1}^1 d x \, e^{-i x \nu} x \Delta g (x) \,  \ . 
 \label{glPDF}
\end{align}
Thus, to extract $x\Delta g (x)$,   we should choose the operators with particular  combinations of the 
$\{\mu \alpha; \lambda \beta  \}$ indices  that contain \mbox{$\widetilde{\mathcal{M}}_{ps}^{(+)}$} 
and $ \widetilde{\mathcal{M}}_{pp}$
in their parametrization.  

It is worth stressing  that  it is  the momentum-weighted density  \mbox{$x \Delta g(x) $}  
that is a natural quantity in this  definition of  the polarized gluon  PDF.
Since  $\widetilde{ M} _{+ \alpha;  \beta +}  (z_-,p)$ is an odd function of $z$,  $x \Delta g(x) $ is an odd function of $x$.
Hence, $ {\cal I}_p (\nu)$ is an odd function of $\nu$,
and, for $\nu >0$ it  can be written as a sine transform
 \begin{align}
{\cal I}_p (\nu)  = 
 \int_{0}^1 d x \, \sin(x \nu) \,  x \Delta g (x) \,  \ . 
 \label{glPDFsin}
\end{align}

An important quantity  is the spin $\Delta G$ contributed by the gluons to the total nucleon spin.
It is given by the integral of $\Delta g(x)$ over all  positive  $x$.
As noted in \mbox{Ref. \cite{Braun:1994jq}}, this integral may  also be written
as an integral over the Ioffe-time distribution 
 \begin{align}
\Delta G  \equiv  \int_{0}^1 d x \, \Delta g (x) =
\int_0^\infty d \nu \, {\cal I}_p (\nu)  
\,  \ . 
 \label{glPDF}
\end{align}
Thus, to estimate  $\Delta G$,  it is sufficient  to know the 
Ioffe-time distribution ${\cal I}_p (\nu)$, without converting it  into the  PDF $ \Delta G (x)$.

\subsection{Matrix elements for extraction of $\Delta g(x)$ }

Since the gluon tensor $G_{\rho \sigma}$  is antisymmetric  with  respect to its indices,
the values $\alpha=+$ and $\beta= +$ may  be taken off the summation in Eq. (\ref{lcpp}). 
Furthermore, since $g_{- -} =0$, the combination  
$g^{\alpha \beta}  \widetilde{ M}_{+ \alpha; \beta +}  (z,p)$ involves the summation over  the 
transverse indices $i,j =1,2$  only, i.e. it reduces 
to $g^{ij}  \widetilde{ M}_{+ i;  j +}  (z,p) \equiv  \widetilde{ M}_{+ i;  + i}  (z,p)  $ (summation over $i$  implied),
for which   we have 
  \begin{align}
 \widetilde{ M}_{+ i;  +i}  
 =   \widetilde{ M}_{0 i;  0i}  +    \widetilde{ M}_{3 i;   3i}  +   \widetilde{ M}_{0 i;  3i}  +
 \widetilde{ M}_{3 i;    0i}   \ . 
\label{ii}
\end{align}  
When $z$ has just the third component, i.e.,  $z=z_3$, the decomposition of these 
combinations  in the basis of   the  $\widetilde{\mathcal{M}}$ structures is
given by 
 \begin{align}
 \widetilde{ M}_{0 i;   0 i}    =  & - 2  s_0  p_0  \widetilde{\mathcal{M}}_{sp} ^{(+)}  +2p_0^2 s_3 z_3   \widetilde{\mathcal{M}}_{pp} 
 +2 s_3 z_3   \widetilde{\mathcal{M}}_{gg}   \ ,
 \label{0i0i}\\
   \widetilde{ M}_{3 i;  3 i  }    = & - 2   p_3 s_3 \widetilde{\mathcal{M}}_{sp} ^{(+)}    -
   2   z_3 s_3   \widetilde{\mathcal{M}}_{sz}^{(+)} \nn &
    + 2s_3 z_3  [ p_3^2  \widetilde{\mathcal{M}}_{pp}- \widetilde{\mathcal{M}}_{gg}+z_3^2  \widetilde{\mathcal{M}}_{zz}+
    z_3 p_3 \widetilde{\mathcal{M}}_{zp}^{(+)}] \ ,
    \label{3i3i}\\
 { M}_{0 i;  3 i  }  = & -2  \left(s_0 p_3   \mathcal{M}_{sp}   + s_3 p_0\mathcal{M}_{ps}   \right)   -2  s_0 z_3   \mathcal{M}_{sz} - 2(sz) \left(    p_0 p_3  \mathcal{M}_{pp}  + p_0 z_3  \mathcal{M}_{pz}  \right) 
  \label{0i3i}   \\ 
 { M}_{3 i;  0 i  }   =  &  -2  \left(s_3 p_0   \mathcal{M}_{sp}   + s_0 p_3\mathcal{M}_{ps}   \right)    -2   s_0 z_3\mathcal{M}_{zs}   -2(sz) \left(     p_3 p_0  \mathcal{M}_{pp}  +  z_3 p_0  \mathcal{M}_{zp}  \right) \ , 
    \label{3i0i}
 \end{align}
where $  \widetilde{\mathcal{M}}_{sz}^{(+)}  =  \widetilde{\mathcal{M}}_{sz} +\widetilde{\mathcal{M}}_{zs}    $, etc.

One may be  tempted  to get the ``light-cone combination'' ${ \widetilde{\mathcal{M}}}_{ps}^{(+)}-
 \nu  \widetilde{\mathcal{M}}_{pp}$ by 
adding these three projections like in Eq. (\ref{ii}). The result (for $z=z_3$) is   given by 
    \begin{align}
    \widetilde{ M}_{0 i;  0i} & +    \widetilde{ M}_{3 i;   3i}  +   \widetilde{ M}_{0 i;  3i}  +
 \widetilde{ M}_{3 i;    0i}   \nn  =  & - 2  s_+ p_+   \widetilde{\mathcal{M}}_{sp} ^{(+)} 
   +2 s_3 z_3 p_+^2  \widetilde{\mathcal{M}}_{pp} 
  -
   2   s_+  z_3 \widetilde{\mathcal{M}}_{sz}^{(+)} +2s_3 z_3^3  \widetilde{\mathcal{M}}_{zz}
    + 2s_3 z_3^2   p_+\widetilde{\mathcal{M}}_{zp}^{(+)} 
 \ , 
\label{all}
\end{align}  
where $p_+=p_0+p_3$ and $s_+=s_0+s_3$. 

One can see that only  the  first two terms  on the right hand side resemble
the combination that  we had in the case of  a  light-cone separation. 
The other terms are   built from  the contaminating ``Euclidean'' terms,
which are  completely absent in the expression
(\ref{lcpp})  for  the $z=z_-$ function 
$ g^{\alpha \beta} \widetilde{ M}_{+ \alpha;  \beta  +}  (z_-,p) $.  

Looking at the  projection  $\widetilde{ M}_{0 i;  0i}  $ (\ref{0i0i}), we see that it  is rather   close    in  structure 
to the desired combination $\widetilde{\mathcal{M}}_{ps}^{(+)}-
 \nu  \widetilde{\mathcal{M}}_{pp}$. Still,   $\widetilde{ M}_{0 i;  0i}  $  contains the $ \widetilde{\mathcal{M}}_{gg}$ contamination.
 Fortunately,  this term can  be subtracted  if we notice that
\begin{align}
 \widetilde{ M}_{ij;   ij}   =  & - 2 s_3 z_3   \widetilde{\mathcal{M}}_{gg}  \  .
  \label{ij} 
  \end{align}
  This observation suggests  to  arrange the combination 
    \begin{align}
 \widetilde{ M}_{0 i;   0 i} + \widetilde{ M}_{ij;   ij}    =  & - 2  s_0  p_0  \widetilde{\mathcal{M}}_{sp} ^{(+)}  +2p_0^2 s_3 z_3   \widetilde{\mathcal{M}}_{pp} 
 \label{00m}
     \end{align}
     that contains just  $\widetilde{\mathcal{M}}_{sp} ^{(+)}  $ and $  \widetilde{\mathcal{M}}_{pp} $.

 Taking \mbox{$p= \{p_0, 0_\perp, p_3 \}$,}  using the requirement \mbox{$(sp)=0$}
 and the normalization condition $s^2=-m^2$, we get $s= \{p_3, 0_\perp, p_0 \}$ 
for the polarization vector in the direction of  the momentum.
This gives
  \begin{align}
 \widetilde{ M}_{0 i;   0 i}  + \widetilde{ M}_{ij;   ij}  =  & - 2  p_3 p_0  \widetilde{\mathcal{M}}_{sp}^{(+)}  +2p_0^3 z_3   \widetilde{\mathcal{M}}_{pp}   \  . 
  \label{0i0iA} 
  \end{align}
  Rewriting the right-hand side as     
\begin{align}
 \widetilde{ M}_{0 i;   0 i}  + \widetilde{ M}_{ij;   ij} = 
  & - 2  p_3 p_0 \left  [ \widetilde{\mathcal{M}}_{sp}^{(+)}- \nu    \widetilde{\mathcal{M}}_{pp}   - \frac{m^2}{p_3^2}  \nu    \widetilde{\mathcal{M}}_{pp}\right ]  \  , 
 \end{align}
 we see that this combination becomes proportional  to the  desired amplitude 
 \mbox{$\widetilde{\mathcal{M}}_{sp}^{(+)}  - \nu    \widetilde{\mathcal{M}}_{pp} $} 
 for large $p_3$. The $p_3^2$-dependence of the remaining term may be used to separate 
  \mbox{$\widetilde{\mathcal{M}}_{sp}^{(+)}  - \nu    \widetilde{\mathcal{M}}_{pp} $} and $({m^2}/{p_3^2})  \nu    \widetilde{\mathcal{M}}_{pp}$, 
 thus  extracting  \mbox{$\widetilde{\mathcal{M}}_{sp}^{(+)}  - \nu    \widetilde{\mathcal{M}}_{pp} $}.
Alternatively, writing the ratio
 \begin{align}
- &\left [\widetilde{ M}_{0 i;   0 i}  + \widetilde{ M}_{ij;   ij} \right ]/(2  p_3 p_0) 
=
  \left  [ \widetilde{\mathcal{M}}_{sp}^{(+)}  -\nu \widetilde{\mathcal{M}}_{pp} \right ]  -\frac{m^2z_3^2 }{\nu}   \widetilde{\mathcal{M}}_{pp}   \  
    \label{00+ii}
 \end{align}
in terms of $\nu$ and $z_3^2$ variables, one may hope to pick out  
$\widetilde{\mathcal{M}}_{sp}^{(+)}  -\nu \widetilde{\mathcal{M}}_{pp} $  exploiting  the strong extra 
$z_3^2$ dependence of the remaining term.
 
 In a similar way, the $ \widetilde{\mathcal{M}}_{gg}$ term  may be excluded from  $\widetilde{ M}_{3 i;  3 i  } $ 
 (\ref{3i3i}) by 
 building  the projection 
 \begin{align}
   &\widetilde{ M}_{3 i;  3 i  } - \widetilde{ M}_{ij;   ij}   =  - 2   p_3 p_0 [  \widetilde{\mathcal{M}}_{sp}^{(+)}    - \nu  \widetilde{\mathcal{M}}_{pp} ] \nn & 
    -
   2   z_3 p_0 \widetilde{\mathcal{M}}_{sz} ^{(+)}  +2   p_0 z_3^3  \widetilde{\mathcal{M}}_{zz}  
 +2   p_0  p_3   z_3^2   \widetilde{\mathcal{M}}_{pz}^{(+)}  \ . 
 \label{3i3iA}
    \end{align}
  Note that it   contains $ \widetilde{\mathcal{M}}_{sp}^{(+)}  $ and $ \widetilde{\mathcal{M}}_{pp}$  in exactly the desired combination.
    Still, there   remain three  contaminations. 
 As  they all come with $z_3$ factors, one may hope that  these terms  are  suppressed  for small $z_3$.
 
 Finally, the remaining  projections (\ref{0i3i}), (\ref{3i0i})   
   \begin{align}
  { M}_{0 i;  3 i  }  = & -2p_0^2 \left ( \widetilde{\mathcal{M}}_{sp} ^{(+)}- \nu \widetilde {\mathcal M}_{pp}  \right )  +2m^2 \widetilde{\mathcal{M}}_{sp}   -2  \nu   \mathcal{M}_{sz} 
  + 2  p_0^2 z_3^2  \mathcal{M}_{pz}  \ , 
  \label{0i3iA}   \\ 
 { M}_{3 i;  0 i  }   =  &  -2p_0^2 \left ( \widetilde{\mathcal{M}}_{sp} ^{(+)}- \nu \widetilde{\mathcal M}_{pp}  \right ) +2  m^2 \widetilde{\mathcal{M}}_{ps}  -2   \nu \mathcal{M}_{zs}  
  +2  p_0^2 z_3^2  \mathcal{M}_{zp}  \ , 
    \label{3i0iA}
    \end{align}
 contain, again,  $ \widetilde{\mathcal{M}}_{sp}^{(+)}  $ and $ \widetilde{\mathcal{M}}_{pp}$ in the combination 
$-2 p_0^2[\widetilde{\mathcal{M}}_{sp}^{(+)}    - \nu  \widetilde{\mathcal{M}}_{pp} ] $ plus $ 2m^2\widetilde{\mathcal{M}}_{sp} $
or $ 2m^2\widetilde{\mathcal{M}}_{ps} $. 
Hence, they are 
 proportional to  $\widetilde{\mathcal{M}}_{ps}^{(+)}-
 \nu  \widetilde{\mathcal{M}}_{pp}$  for large $p_0$, but  have two other contaminations. 
 
  A possible    advantage of  $\widetilde{ M}_{0 i;  3 i  } $ and $  \widetilde{ M}_{3 i;  0 i  } $ is that   they have 
 $ 2p_0^2  $  factor in front of $ \widetilde{\mathcal{M}}_{sp}^{(+)} $,
 while  we have  the $2p_3 p_0$ factor in the case of $\widetilde{ M}_{0 i;   0 i}  + \widetilde{ M}_{ij;   ij} $.
  Hence,   $\widetilde{ M}_{0 i;  3 i  } $ and $  \widetilde{ M}_{3 i;  0 i  } $   may    have a stronger 
  signal  for small $p_3$
 than  \mbox{$ \widetilde{ M}_{0 i;   0 i}  + \widetilde{ M}_{ij;   ij}$.}

 \subsection{Relation to ${\bf E}$ and ${\bf B}$ fields}
  \label{2.5}
  
  So far, our  parametrization was based on  the most  general 
  properties   of  matrix elements, like Lorentz covariance and antisymmetry 
  of  $G_{\rho \sigma}$ with  respect to its indices.
Now, let us  incorporate the  fact  that we deal with the matrix element
$G(z) \widetilde G(0)$ in which both $G$ and $\widetilde G$ may be written in terms 
of the electric $E_k$ and magnetic $B_k$ fields.

Namely, we have 
$ G_{0i} = E_i$,  $ \widetilde{G}_{0i } =B_i $,
$
G_{ij}=- \epsilon_{ijk} B_k \  , 
$
\mbox{$
 \widetilde{G}_{i j} =  \epsilon_{ijk} E_k $,}  with  the familiar ${\bf E} \leftrightarrow {\bf B}$ interchange 
when $G \to \widetilde G$.  
To treat the fields in a more symmetric way, we use  translation invariance 
of the forward matrix elements, and  shift the arguments of the fields by $z/2$ to  find 
  \begin{align}
 &\widetilde{ M}_{0 i;   0 i}  \left (z\right ) =   \left \langle E_{i}\left (z/2\right ) B_{i}\left (-z/2\right )\right \rangle  - \{z \to -z \} \nn &
 = \left \langle {\bf E}_\perp \left (z/2\right ) \cdot  {\bf B}_\perp \left (-z/2\right )\right \rangle
  - \{z \to -z \}
  \end{align}
  and 
  \begin{align}
   \widetilde{ M}_{3 i;  3 i  } \left (z\right )   =   & - \Bigl [\left \langle \epsilon_{3ik}B_k \left (z/2\right ) \epsilon_{3il}  E_{l}\left (-z/2
   \right ) \right 
   \rangle
    - \{z \to -z \} 
   \Bigr ]\nn & =
    -\Bigl [  \left \langle B_{k}\left (z/2\right )E_{k} \left (-z/2\right ) \right \rangle     - \{z \to -z \} \Bigr ] \nn      &=  \widetilde{ M}_{0 i;   0 i} \left (z\right ) \ . 
\end{align}
Thus, we arrive at the  relation 
  \begin{align} 
  \widetilde{ M}_{3 i;  3 i  } \left (z\right )  =  \widetilde{ M}_{0 i;  0 i  } \left (z\right )  \  .
  \label{0=3}
    \end{align}
   Basically, it  results from the fact  that changing $0i$ into $3i$ corresponds 
   to the ${\bf E} \leftrightarrow {\bf B}$ interchange, which makes no change in the 
   ${\bf E} \leftrightarrow {\bf B}$-symmetric $G \widetilde G$ operator.

     However,  Eq. (\ref{0=3}) 
   looks  rather unexpected in view of different structure of 
  the decompositions 
   (\ref{0i0i}) and  (\ref{3i3i}) for these projections. Combining these decompositions with  
   \mbox{Eq. (\ref{0=3})}   results in 
 the ``sum rule''
        \begin{align}
  & 
       \  2    \widetilde{\mathcal{M}}_{gg} =-\widetilde{\mathcal{M}}_{zs} ^{(+)}  -m^2    \widetilde{\mathcal{M}}_{pp}       + z_3^2  \widetilde{\mathcal{M}}_{zz}  
 +    \nu    \widetilde{\mathcal{M}}_{zp}^{(+)}  
  \label{SR} 
    \end{align} 
  involving the invariant amplitudes both from $ \widetilde{ M} _{\mu \alpha;   \lambda  \beta}^{(1)}  (z,p)$
(\ref{M1})   and $ \widetilde{ M} _{\mu \alpha;   \lambda  \beta}^{(2)}  (z,p)$ (\ref{M2}). 
  Substituting this relation for $ \widetilde{\mathcal{M}}_{gg} $ into Eq. (\ref{M2}) 
  changes the tensor coefficients accompanying the invariant amplitudes 
  $\widetilde{\mathcal{M}}_{zs},   \widetilde{\mathcal{M}}_{sz},    \widetilde{\mathcal{M}}_{pp},      \widetilde{\mathcal{M}}_{zz},   \widetilde{\mathcal{M}}_{zp}$ and $  \widetilde{\mathcal{M}}_{pz}$.  As an example, $ \widetilde{\mathcal{M}}_{pp}$ will be accompanied by
  the 
       \begin{align}
& g_{\mu\lambda} \left (p_\alpha p_\beta  - \frac{p^2}{4} g_{\alpha\beta}\right )- g_{\mu\beta} 
\left (p_\alpha p_\lambda  - \frac{p^2}{4} g_{\alpha\lambda}\right ) 
\nn & 
- g_{\alpha\lambda} \left (p_\mu p_\beta  - \frac{p^2}{4} g_{\mu\beta}\right ) + g_{\alpha\beta} \left (p_\mu p_\lambda  - \frac{p^2}{4} g_{\mu\lambda}\right )  
      \end{align}
      factor, in which the original $p^\rho p^\sigma$-type  tensors are substituted  by their traceless versions. 
      The changes to traceless versions  will occur in the structures accompanying 
      all other invariant amplitudes listed above. 
      Another sum rule is derived by considering 
   \begin{align}
 \widetilde{ M}_{ij;   ij}  \left (z\right ) =&   - \Bigl [\left \langle \epsilon_{ijk}B_k \left (z/2\right ) \epsilon_{ijl} E_{l}\left (-z/2
   \right ) \right 
   \rangle
    - \{z \to -z \} 
   \Bigr ]\nn & =
    - 2\Bigl [  \left \langle B_{3}\left (z/2\right )E_{3} \left (-z/2\right ) \right \rangle     - \{z \to -z \} \Bigr ]  
    \nn      &= 2 \widetilde{ M}_{0 3;   0 3} \left (z\right )  \  .
        \label{ij=03}
    \end{align}
    Thus, we have 
      $
   \widetilde{M}_{ij;   ij}  \left (z\right ) =  2 \left \langle {\bf E}_3 \left (z/2\right ) \cdot  {\bf B}_3 \left (-z/2\right )\right \rangle
  - \{z \to -z \} \ .
  $
    To use the resulting relation
      $
 \widetilde{ M}_{ij;   ij}  \left (z\right ) = 2 \widetilde{ M}_{0 3;   0 3} \left (z\right )  \  , 
 $
    we need the  decomposition 
      \begin{align}
 \widetilde{ M}_{0 3;   0 3}=& 
   p_{0} z_{3} (p_0 s_{3}-p_3 s_0)  \widetilde{\mathcal{M}}_{p s p z}^{(+)}  
    + s_0 p_0 z_3^2
  \widetilde{\mathcal{M}}_{s z p z}^{(+)} - s_3p_0^2 z_3^3\widetilde{\mathcal{M}}_{p p z z}
 \nn  & +s_{3} z_{3} 
  \left(\widetilde{\mathcal{M}}_{s z}^{(+)} +m^{2} \widetilde{\mathcal{M}}_{p p}-z_{3}^{2} \widetilde{\mathcal{M}}_{z z}-\nu \widetilde{\mathcal{M}}_{z p}^{(+)} +\widetilde{\mathcal{M}}_{gg} \right) , 
\end{align}
where $\widetilde{\mathcal{M}}_{p s p z}^{(+)}=\widetilde{\mathcal{M}}_{p s p z}+\widetilde{\mathcal{M}}_{p z p s}$,  and, similarly,  
\mbox{$\widetilde{\mathcal{M}}_{s z p z}^{(+)}=\widetilde{\mathcal{M}}_{s z p z}+\widetilde{\mathcal{M}}_{p z s z}$} .
Using the sum rule  (\ref{SR}) simplifies this expression into 
   \begin{align}
 &\widetilde{ M}_{0 3;   0 3}= 
   p_{0} z_{3} (p_0 s_{3}-p_3 s_0)  \widetilde{\mathcal{M}}_{p s p z}^{(+)} 
  + s_0 p_0 z_3^2
  \widetilde{\mathcal{M}}_{s z p z}^{(+)} - s_3p_0^2 z_3^3\widetilde{\mathcal{M}}_{p p z z}-s_{3} z_{3} 
 \widetilde{\mathcal{M}}_{gg} \ .  
\end{align}
    Applying now  $ \widetilde{ M}_{0 3;   0 3}=\frac12 \widetilde{ M}_{ij;   ij}= -  s_3  z_3   \widetilde{\mathcal{M}}_{gg}  $, we obtain 
    the second sum rule
         \begin{align}
s_3p_0 z_3^2\widetilde{\mathcal{M}}_{p p z z}=& (p_0 s_{3}-p_3 s_0)  \widetilde{\mathcal{M}}_{p s p z}^{(+)}  
  + s_0  z_3
  \widetilde{\mathcal{M}}_{s z p z}^{(+)}
 \label{SR2}
\end{align} 
relating the invariant amplitude $\widetilde{\mathcal{M}}_{p p z z}$ from 
$ \widetilde{ M} _{\mu \alpha;   \lambda  \beta}^{(2)} $ with the invariant amplitudes 
$\widetilde{\mathcal{M}}_{p s p z}^{(+)}  $ and $  \widetilde{\mathcal{M}}_{s z p z}^{(+)}
$ from $ \widetilde{ M} _{\mu \alpha;   \lambda  \beta}^{(1)} $.

One may ask if there are other relations following from the ${\bf E} \leftrightarrow {\bf B}$ interchange 
symmetries of the $  \widetilde{ M} _{\mu \alpha;   \lambda  \beta}$ matrix element.
To this end, let us list  various possibilities for the set of  indices $\{ \mu \alpha;   \lambda  \beta \}$.
The index $\alpha$ from the first pair may correspond  to 0, 3 or one of the   transverse components 1,2, call it $i$.
Note now that,  on the right-hand sides of the decompositions (\ref{M1}), (\ref{M2}), the index $\alpha$ may be carried by $p_\alpha$, $z_\alpha$
or $s_\alpha$, none of which has transverse components. Hence, if $\alpha=i$, it appears on the right-hand side through
the metric tensor $g_{\alpha \lambda}$ or  $g_{\alpha \beta}$. 
Thus, the matrix element in this case has the structure $\widetilde{ M} _{\mu i ;\lambda i }$
where $i=1$ or $i=2$.  Since $g_{11}=g_{22}$, we conclude that 
$\widetilde{ M} _{\mu 1 ; \lambda1 }=\widetilde{ M} _{\mu 2 ; \lambda 2 }$.
This means that, without a loss of generality, we can consider the sum $\sum_{i=1}^2 \widetilde{ M} _{\mu i ; \lambda i }$,
which from now on we will denote simply as  $\widetilde{ M} _{\mu i ; \lambda i }$,
implying summation over $i$,  just as we did before.

For the  remaining indices $\mu, \lambda$,  we have 5 possibilities: $\{\mu, \lambda\} = \{0,0 \},  \{3,3 \},  \{j,j \}$, and $ \{ 0,3\},   \{3,0 \}$. 
We have already obtained the relations involving the first three possibilities, namely
$  \widetilde{ M}_{0 i;  0 i  } =   \widetilde{ M}_{3 i;  3 i  } \left (z\right )  $ and 
$ \widetilde{ M}_{ji;   ji}  \left (z\right ) = 2 \widetilde{ M}_{0 3;   0 3} \left (z\right )  $. 
The second relation, in fact, covers the situation when neither of indices $\mu$ and $ \alpha$ 
of the first pair 
is  transverse. 

The remaining cases correspond to $  \widetilde{ M}_{0 i;  3 i  }  $ and $  \widetilde{ M}_{3 i;  0 i  } $. 
Let us write  the relevant bilocal operators in terms of ${\bf E}$ and ${\bf B}$ fields. For the first of them, we have 
 \begin{align}
\widetilde{ M}_{0 i;  3i} \left (z\right )=  G_{01} \left (z/2\right ) G_{02} \left (-z/2\right )- G_{02}\left (z/2\right ) G_{01}\left (-z/2\right )
 - \{z \to -z \}  \  . 
\end{align}
Hence, this  matrix element involves just the electric field 
 \begin{align}
 \widetilde{ M}_{0 i;  3i} \left (z\right )=2 \left \langle {\bf E}_\perp \left (z/2\right ) \times {\bf E}_\perp \left (-z/2\right )\right \rangle _3 \,  , 
 \end{align}
bringing in no restrictions on invariant amplitudes. Similarly, the matrix element 
\begin{align}
 \widetilde{ M}_{3 i;  0i} \left (z\right )=  G_{31} \left (z/2\right ) G_{23} \left (-z/2\right )- G_{32}\left (z/2\right ) G_{13}\left (-z/2\right )
 - \{z \to -z \} \  
\end{align}
is built from the operator containing  the magnetic field only
\begin{align}
  \widetilde{M}_{3 i;  0i} \left (z\right )= -2 \left \langle {\bf B}_\perp \left (z/2\right ) \times {\bf B}_\perp \left (-z/2\right )\right \rangle _3 \ ,
 \end{align}
thus  producing no extra restrictions on  invariant amplitudes. 

\subsection{Multiplicatively renormalizable combinations} 
\label{mult}

Off the light cone, the $ \widetilde { M}_{\mu \alpha;   \lambda  \beta}$  matrix elements have extra  ultraviolet divergences
related to presence of the gauge link.
 For any   set of its indices $\{ \mu \alpha;  \lambda \beta  \}$, each matrix element 
 is multiplicatively renormalizable with respect to these divergences \cite{Li:2018tpe}.
 However,  
 in general, the 
 anomalous dimensions  are  different.

In Ref. \cite{Zhang:2018diq}, it was established that  the combinations  represented in Eq. (\ref{ii}),  namely,  
$  \widetilde { M}_{0 i; i 0 }  $,    $  \widetilde { M}_{3 i;  i 3 } $,    \mbox{${ \widetilde  M}_{0 i; i3 }  ,
 {  \widetilde M}_{3 i; i 0 }  $},  with   
 summation over     transverse indices $i$,  
 are each  multiplicatively renormalizable at the one-loop level. 
 Furthermore,   as  we will see,  the combination $G_{ij}\widetilde G_{ij}$  (with summation over  transverse $i,j$) 
  has the same one-loop UV anomalous dimension
 as  ${\widetilde  M}_{0 i;  i 0 } $,  while the matrix element of
  $G_{30}\widetilde G_{03}$ has the same one-loop UV anomalous dimension as ${M}_{3 i;  i 3 } $. 
 Hence,    the combinations of  Eqs. (\ref{00m}) and (\ref{3i3iA}) are 
 multiplicatively renormalizable at the one-loop level.

\subsection{Reduced Ioffe-time distribution}

Within the pseudo-PDF approach \cite{Radyushkin:2017cyf}, the 
link-related UV divergences are eliminated through introducing
the reduced Ioffe-time distribution.  Namely, for each  multiplicatively  renormalizable amplitude $  \mathcal{M}$  we build the ratio 
 \begin{align}
{\mathfrak M} (\nu, z_3^2) \equiv \frac{{\mathcal{M}} (\nu, z_3^2)}{{\mathcal{M}}(0, z_3^2)} \  , 
 \label{redm0}
\end{align}
in which 
the   link-related   UV divergent $Z(z_3^2 \mu^2_{UV} )$  factors generated by the
vertex 
 and  link  self-energy diagrams 
cancel.  As a result, the small-$z_3^2$ dependence of the reduced 
pseudo-ITD  ${\mathfrak M} (\nu, z_3^2) $  comes from the logarithmic DGLAP evolution
effects only. 


 \setcounter{equation}{0}  

\section{One-loop corrections}


Our next goal is to develop   one-loop matching relations  for the matrix elements that
may be used in the lattice extraction of the polarized gluon PDF.
In their calculation, we have  used  the same method  
 \cite{Balitsky:1987bk} that was used in Refs. \cite{Balitsky:2019krf, Balitsky:2021qsr} 
 for the unpolarized case.

\subsection{Link self-energy  contribution}

\begin{figure}[h]
   \centerline{\includegraphics[width=2in]{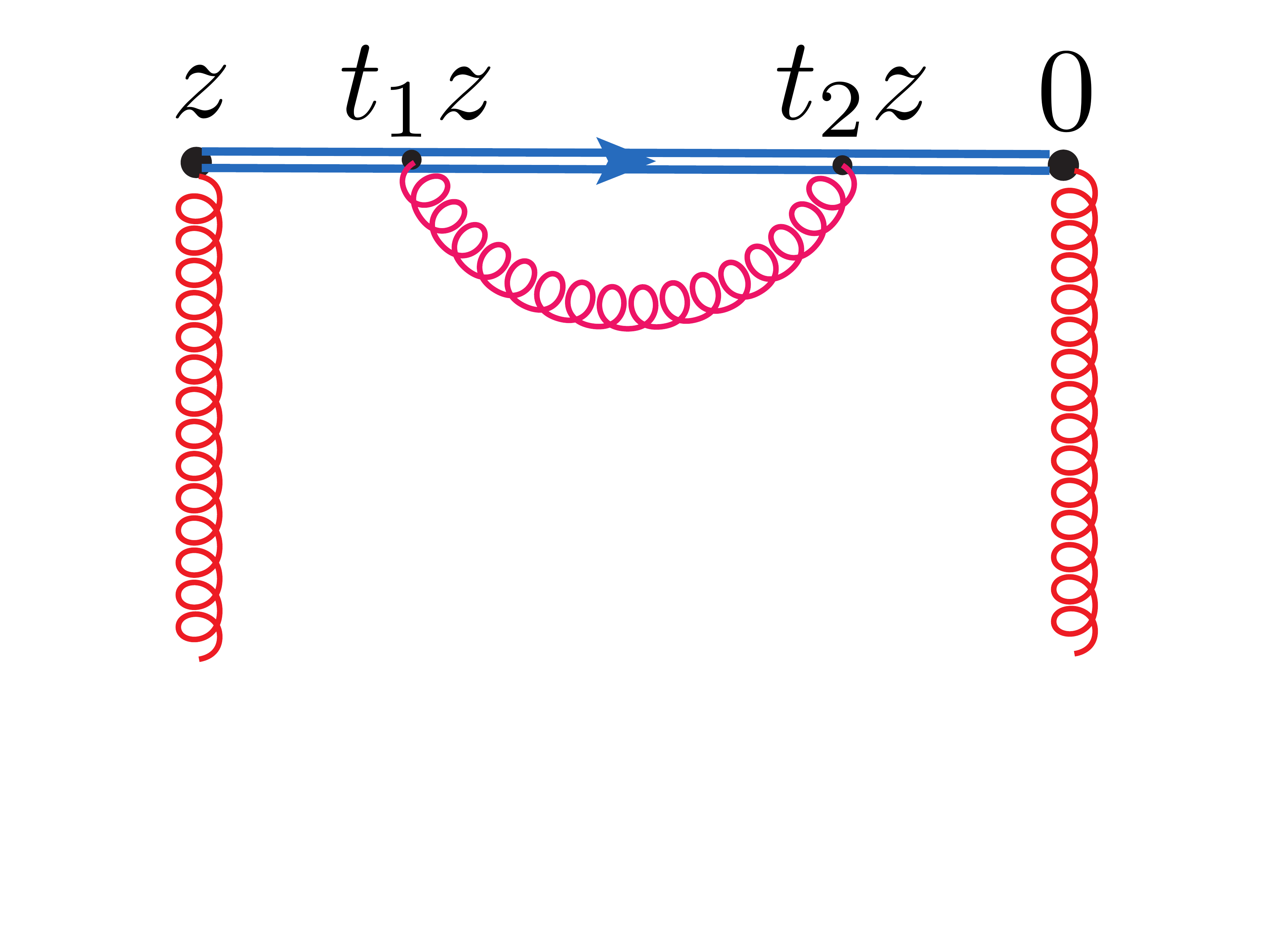}}
        \vspace{-8mm}
   \caption{Self-energy-type  correction for  the gauge link.
   \label{linkself}}
   \end{figure}

  The 
self-energy correction  for  the  gauge 
 link is given by the simplest diagram  (see Fig. \ref{linkself}). 
  In  lattice perturbation theory, it was 
  calculated 
at one loop in Ref. \cite{Chen:2016fxx}. 
 The result is close to that given by the  expression
  \begin{align} 
 \Gamma_{\rm UV}(z_3,a) 
\sim   - \,\frac{\alpha_s}{2\pi}\, N_c
    \left [  
   \,
 2  \frac{ |z_3|}{a} \,  \tan
   ^{-1}\left(\frac{|z_3|}{a}\right)    -   \,  \ln 
   \left(1+ \frac{z_3^2}{a^2}\right)  \right ] 
   \label{linkse}
 \end{align}
 obtained using 
Polyakov regularization $1/z^2 \to 1/(z^2-a^2)$ 
 for  the gluon 
propagator in the coordinate space, with the parameter $a$ related to
the lattice spacing by $a=a_L/\pi$. 
An important property of this contribution is the presence of 
  a $\sim z_3/a_L$ linear term,
  where $a_L$ is  the lattice spacing 
  that provides here the   ultraviolet cut-off.

  Clearly, this  correction is  just a function of $z_3$. 
  It does not induce  any  $\nu$-dependence, and the resulting $\nu$-independent  factors   cancel in the ratio (\ref{redm0}).
  For this reason, 
   the  explicit form of this factor is  not very essential in the pseudo-PDF approach. 
   
For completeness, we present here the  expression for the link self-energy digram in
Feynman gauge obtained using the  dimensional regularization, 
 \begin{align}
&
-{g^2N_c \over 4\pi^2[(-z^2 \mu_{\rm UV}^2 +i\epsilon)]^{{d\over 2}-2}} {\Gamma\big({d/ 2}-1\big) \over (3-d)(4-d)}
G_{\mu\alpha}(z)G_{\lambda \beta }(0) \ ,
\label{selfAD}
\end{align}
where the pole for $d=3$ ($d=4$) corresponds to the linear (logarithmic)  UV divergences
present in this diagram.

   \subsection{UV divergent vertex terms} 
   
\begin{figure}[h]
   \centerline{\includegraphics[width=2.5in]{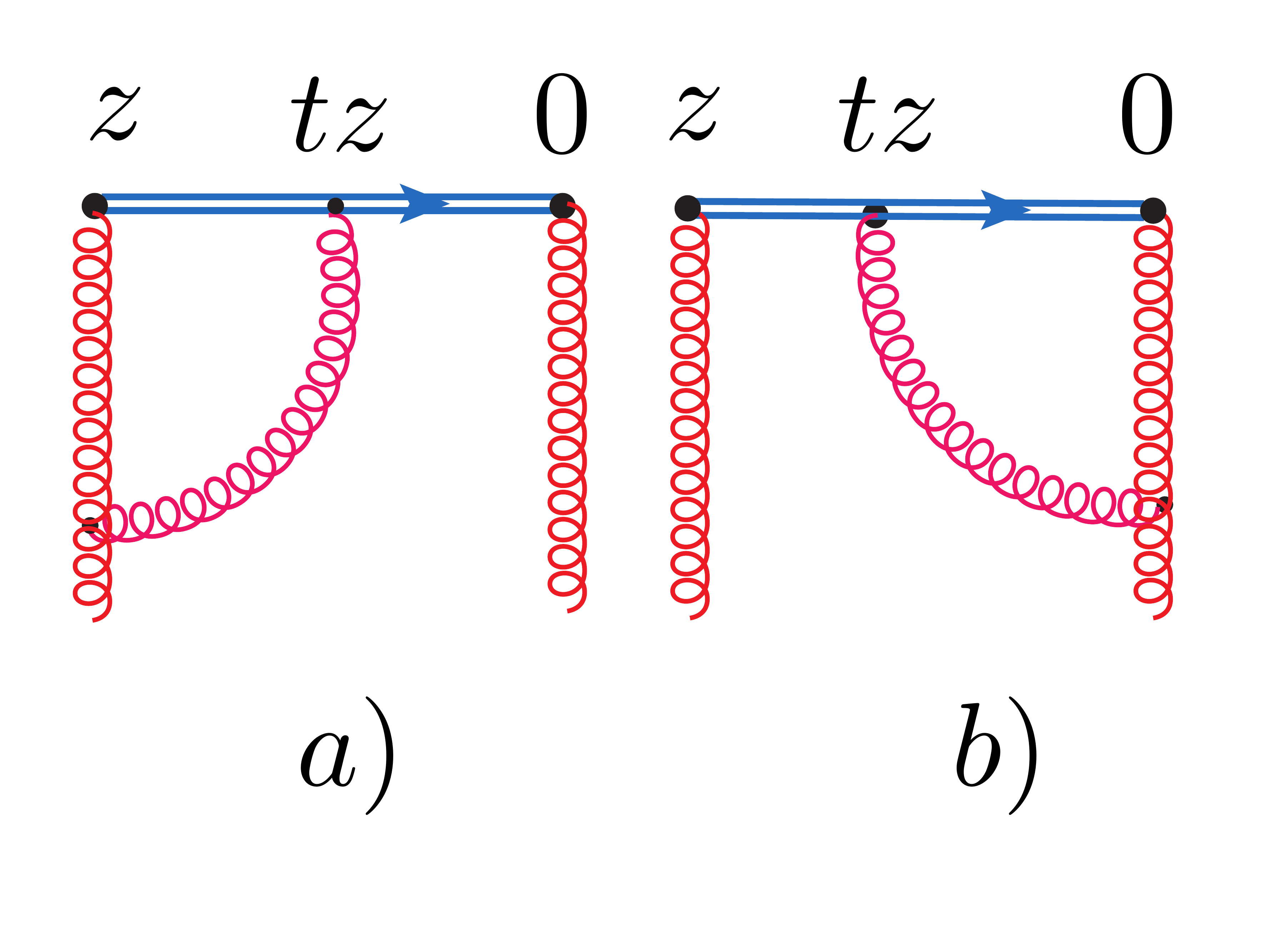}}
        \vspace{-8mm}
   \caption{Vertex diagrams with  gluons coming out of the gauge link.
   \label{link}}
   \end{figure}

    UV divergent terms are also present in     vertex diagrams     
involving gluons  that connect the gauge link with the gluon  lines, see  \mbox{Fig. \ref{link}. } 
Clearly, the gluon exchange produces a correction just to one of the fields in the 
$G_{\mu\alpha}(z) \widetilde G_{\lambda \beta }(0)$ operator, while another remains intact. 
A minor complication compared to Refs.  \cite{Balitsky:2019krf, Balitsky:2021qsr}  is 
the presence of a dual field  $\widetilde G$ in one of the vertices. But this changes  only the tensor structure  of the contributions
without affecting the integral. 
\vspace{-2mm} 

As established in Refs. \cite{Balitsky:2019krf, Balitsky:2021qsr}, the vertex correction 
may be represented as  the sum of  the UV divergent and UV finite    parts. 
   The UV-divergent  part  of the vertex correction to $G_{\mu\alpha}(z)$  is   given by 
\begin{align}
&\frac{N_c g^2}{8\pi^2} 
\frac{\Gamma(d/2-1)}{(d-2 )(-z^2)^{d/2-1}}
\int_0^1 \dd u\,  \left(u^{3-d}-u\right)  
 \left(z_{\alpha} G_{z\mu}( \bar uz) - z_{\mu} G_{z\alpha}(\bar uz)\right) \ , 
 \label{2a1}
 \end{align}
 where $G_{z\sigma} \equiv z^\rho G_{\rho\sigma}$ and $\bar u \equiv 1-u$.  
 As we see, the overall
 \mbox{$d$-dependent}  factor  here is finite for $d=4$,  but  the \mbox{$u$-integral} diverges at the lower limit.
 If one uses the dimensional UV    regularization 
with 
 $d=4-2\varepsilon_{\rm UV}$,  the divergence   converts 
 into a pole at $\euv=0$. 
Isolating the UV  divergence   by taking $\bar u =1$ in the gluonic field produces 
 \begin{align}
&\frac{N_c g^2}{4\pi^2} 
\frac{\Gamma(d/2-1)}{(d-2 )(-z^2)^{d/2-1}}
\left (\frac1{4-d}-\frac12 \right ) 
 \left(z_{\alpha} G_{z\mu}(z) - z_{\mu} G_{z\alpha}(z)\right) \ 
 \label{UVvert}
 \end{align}
 plus the remainder  given by 
 \begin{align}
&\frac{N_c g^2}{8\pi^2} 
\frac{\Gamma(d/2-1)}{(d-2 )(-z^2)^{d/2-1}}
\int_0^1 \dd u\,  \left[u^{3-d}-u\right]_{+(0)}  
 \left(z_{\alpha} G_{z\mu}( \bar uz) - z_{\mu} G_{z\alpha}(\bar uz)\right) \ , 
  \label{UVvertreg}
 \end{align}
 where the plus-prescription at $u=0$  is defined as
   \begin{align}
& \int_0^1  \dd u \left[f(u)\right]_{+(0)} g(u) = \int_0^1  \dd u f(u) [g(u) -g(0)]  \ .
 \label{plus0}
 \end{align}

As  explained in Refs. \cite{Balitsky:2019krf, Balitsky:2021qsr},  if we take $z=z_3$, the 
field ${\cal G}_{\mu\alpha}(z)=
z_{\alpha} G_{z\mu}( z) - z_{\mu} G_{z\alpha}(z)$
 in Eq. (\ref{UVvert}) is actually proportional to 
the field 
$ G_{\mu\alpha}(z)$ in the original operator. 
 In explicit form: 
${\cal G}_{0 i}(z)= 0$,   ${\cal G}_{ij}(z)= 0$,  ${\cal G}_{0 3}(z)= - z_3^2 G_{0 3}( z)$  and   ${\cal G}_{3 i}(z)= - z_3^2 G_{ 3 i }( z) $.
Thus, when one of the 
indices equals 3, 
 we have a nontrivial vertex anomalous dimension (AD, call it $\gamma$),  since 
${\cal G}_{3\alpha}(z)=  - z_3^2 G_{ 3 \alpha }( z) $   for all $\alpha$.
In all other cases, we have a trivial (vanishing)  vertex  AD, since   ${\cal G}_{ij}(z)=0$ and  ${\cal G}_{0i}(z)=0$. 

For the dual field $\widetilde G_{\lambda \beta }$, the ``$\gamma$-counting'' is inverse:
if none of the indices $\lambda,  \beta $ equals 3, the field  has AD   equal to $\gamma$.
Otherwise, its AD is zero.  Combining the ADs from $G$ and $\widetilde G$, we see that the matrix elements 
 $  \widetilde{ M}_{0 i;  0i} $,    $  \widetilde{ M}_{ ij;  ij} $,  $  \widetilde{ M}_{03;  03} $ and  $  \widetilde{ M}_{3 i;  3i} $ all 
   have  vertex AD equal to $ \gamma$;
 while $  \widetilde{ M}_{0 i;  3i} $ has zero AD and $  \widetilde{ M}_{3 i;  0i} $ has AD equal to $2 \gamma$. 
   These observations  lead to  the results announced in Section \ref{mult}. Namely,  the matrix element 
 $\widetilde  M_{ij;ij}$   
  has the same one-loop UV anomalous dimension
 as  ${\widetilde  M}_{0 i;  0i } $,  while   $\widetilde M_{30;03}$ has the same one-loop UV anomalous dimension as ${M}_{3 i; 3 i  } $. 

Of course, the UV cut-off  produced  by the dimensional regularization is 
rather different  from that produced by a finite lattice spacing.
The latter, as pointed out earlier, is similar to the Polyakov regularization   $1/z^2 \to 1/(z^2-a^2)$ 
 for  the gluon 
propagator in the coordinate space, with the parameter $a$ related to
the lattice spacing by $a=a_L/\pi$.  
The UV logarithms $(\alpha_s N_c/4\pi) \ln z_3^2 \mu^2_{\rm UV}$ in this case are substituted by $(\alpha_s N_c/4\pi)\ln (1+z_3^2/a^2)$
(compare with Eq. (\ref{linkse})).
In higher orders, they, as usual,  exponentiate into
   \begin{align}
Z_{\rm L} (z_3/a_L) = \left (1+\pi^2 z_3^2/a_L^2\right )^{\alpha_s N_c/4\pi} \ . 
\label{vAD}
 \end{align}
 For each particular  type of the operator discussed above, one would have $Z^\gamma_{\rm L} (z_3/a_L)$,
 where $\gamma$ is the number (0,  or 1,  or 2) corresponding to the  operator in question.

Building the matching relations for  particular matrix elements entering in the combinations listed  in Eqs. (\ref{00m}),  (\ref{3i3iA})  and  (\ref{3i0iA}),
we will need the following results for  the  UV-divergent parts of  vertex corrections
\begin{align}
&  G_{l i} (z_3) \widetilde G_{li}(0) 
\stackrel{\rm UV}{\longrightarrow}  
 \frac{g^2N_c \Gamma(d/2-1)}{4\pi^2(z_3^2)^{d/2-2}} \int_0^1 \dd u \left(\frac{  u^{3-d}- u}{d-2} \right)   G_{l i}(\bar u z_3)   \widetilde G_{ li}(0)  \ ,
\end{align}
where $l=0, 3$ or $l=j$ (in the latter case, also summation over $j$ is implied).  We also have 
\begin{align}
& G_{3 i} (z_3) \widetilde  G_{0i}(0) 
\stackrel{\rm UV}{\longrightarrow}  
\frac{g^2N_c \Gamma(d/2-1)}{2\pi^2(z_3^2)^{d/2-2}} \int_0^1 \dd u \left(\frac{ u^{3-d}- u}{d-2} \right)G_{3 i}(\bar u z_3)\widetilde G_{0i}(0) 
\end{align}
and 
$
  G_{0 i} (z_3) \widetilde G_{3i}(0) \stackrel{\rm UV}{\longrightarrow}   0$.

   \subsection{Evolution contribution from the vertex diagrams}

The UV finite contribution  from  the vertex diagrams shown in Fig.  \ref{link} generates 
the  evolution $z_3^2$-dependence of the matrix element. 
It   may be symbolically written as  
\begin{align}
&  G_{\mu\alpha}(z_3) \widetilde G_{\lambda \beta }(0)
\stackrel{\rm Evol}{\longrightarrow} 
\frac{g^2N_c \Gamma(d/2-2)}{4\pi^2(z_3^2)^{d/2-2}}  \int_0^1 \dd u  \left[\frac{ u^{3-d}-1}{d-3}\right]_+   G_{\mu\alpha}(\bar u z_3) \widetilde G_{\lambda \beta }(0) \ .
\end{align}
In this case,   the gluonic operator 
has the same tensor structure as the original operator 
$  G_{\mu\alpha}(z_3) \widetilde G_{\lambda \beta }(0)$ 
differing from it just by 
rescaling $z \to \bar u z$. 
 There is no mixing with operators  of a different type. 
 Also, the evolution factor is the same for any combination of the indices in $G_{\mu\alpha} \widetilde G_{\lambda \beta }$.
 
 The $u$-integral  now  does not diverge   for $d=4$, but the  overall 
\mbox{$\Gamma(d/2-2)$}  factor  has a pole  $1/(d-4)$. 
Note that  the singularity 
for $d=3$ from the  pole   $1/(d-3)$ formally corresponds  to a linear UV divergence. 
However, it  is compensated by a zero coming for $d=3$ from the $\left [u^{3-d}-1\right ]$ combination in the integrand. 
The remaining  $1/(d-4)$  pole  
corresponds to a collinear divergence 
that appears  
 because  all  the propagators and external lines correspond to massless particles.
 The integrand factor $\left [u^{3-d}-1\right ]_+$ for $d=4$ produces the $\left [\bar u/u\right ]_+$
 part of the evolution kernel.

\subsection{Gluon self-energy diagrams}

  \begin{figure}[h]
   \centerline{\includegraphics[width=1.5in]{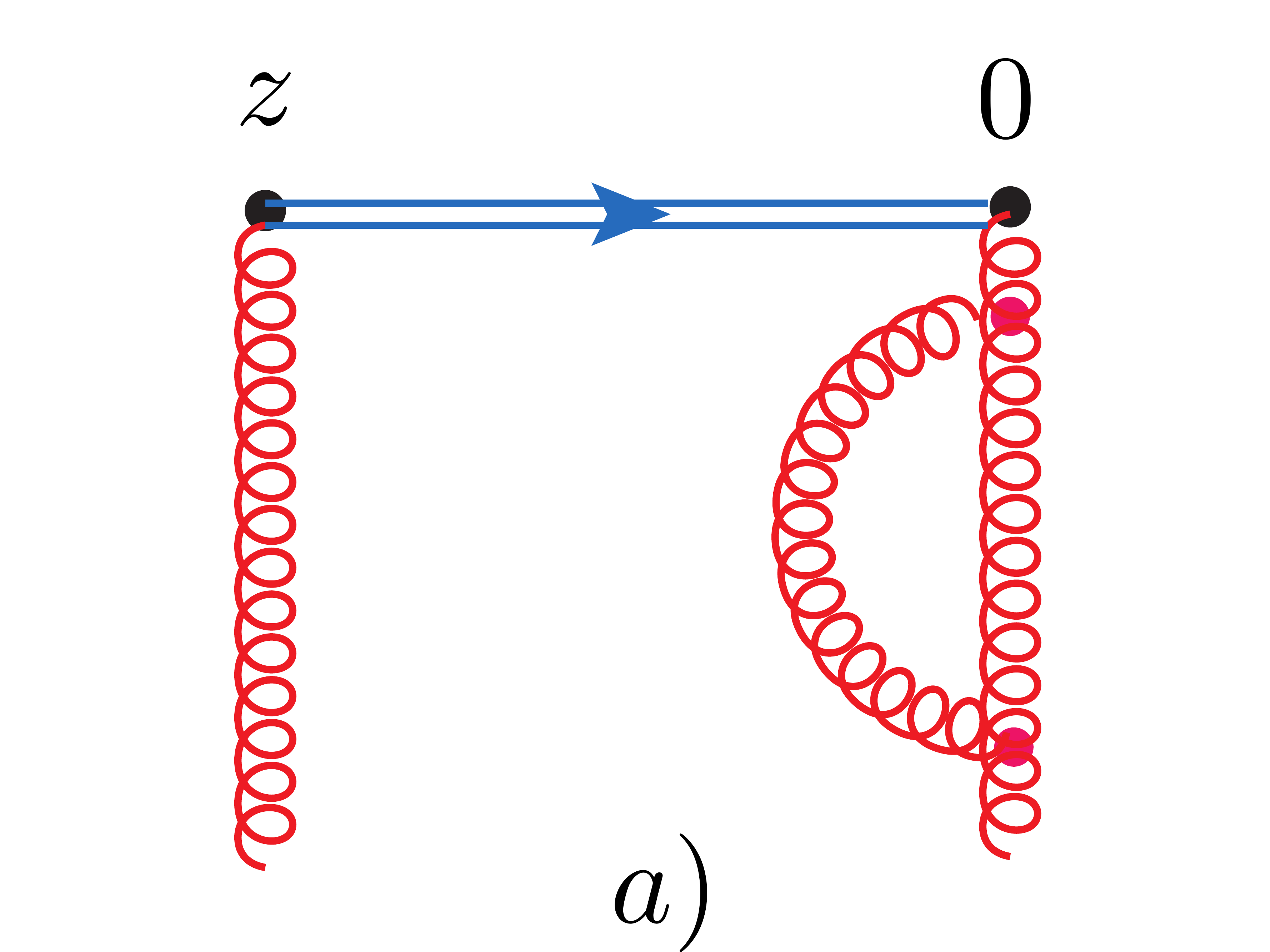} \hspace{-5mm} \includegraphics[width=1.5in]{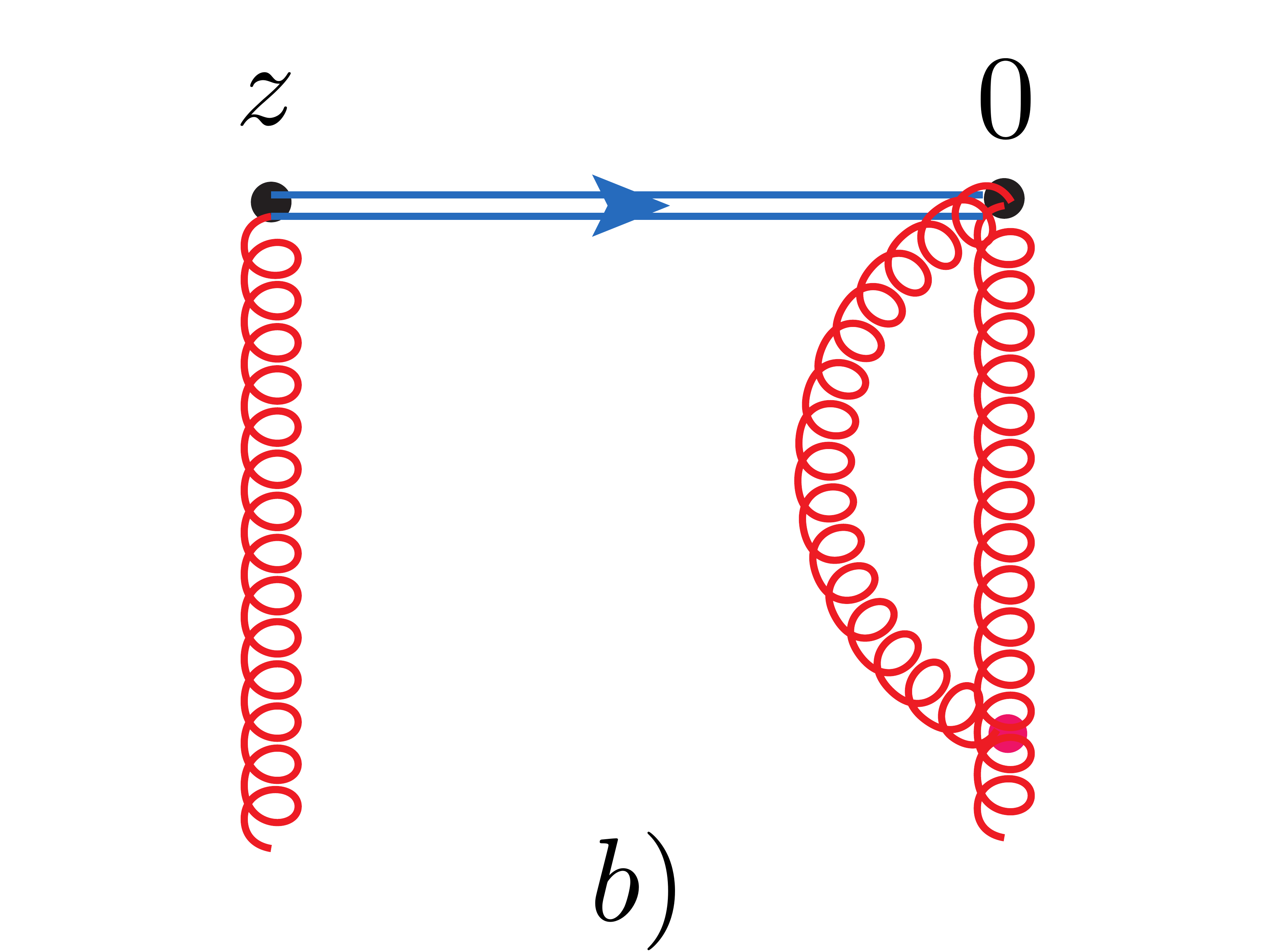}}
   \caption  { Gluon  self-energy-type insertions into the right leg.  
   \label{gluself}}
   \end{figure}

  Another simple type of one-loop corrections is represented by the gluon  self-energy diagrams,  one of which is shown in Fig. \ref{gluself}a.
These diagrams have both the UV and collinear 
divergences. 
The combined contribution 
of the Fig. \ref{gluself} diagrams and their left-leg analogs  is given by
\begin{align}
{g^2N_c\over 8\pi^2} 
\frac1{2-d/2}
\left [2 - \frac{\beta_0}{2N_c} \right ]G_{\mu\alpha}(z)G_{\lambda \beta }(0) \  ,
\label{counter3}
\end{align}
where $\beta_0 =11N_c/3$ in gluodynamics, so that the terms in the square bracket combine into 1/6.

\subsection{Box diagram}

\begin{figure}[h]
   \centerline{\includegraphics[width=1.7in]{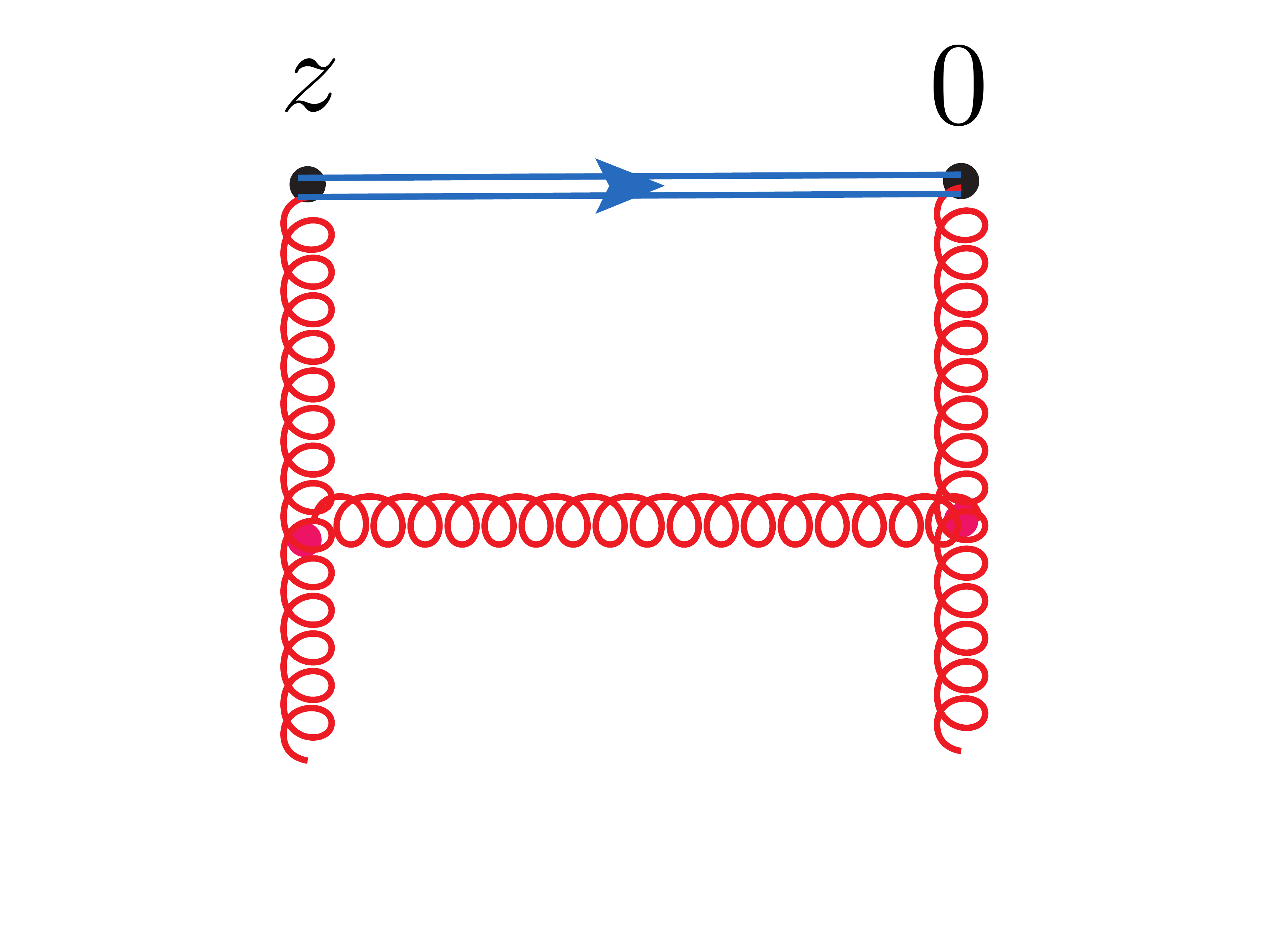}}
        \vspace{-5mm}
   \caption  {Box diagram.     \label{box}}
   \end{figure}

The most nontrivial is the calculation 
  of  the ``box'' diagram  corresponding to 
  a gluon exchange between
two gluon lines (see Fig.   \ref{box}).  
While this diagram has no  UV divergences,  it contains  DGLAP 
$\log z_3^2$ evolution contributions.  
In distinction  to the vertex diagrams, the original  
$G_{\mu\alpha}( z)  G_{\nu \beta }( 0)$ operator generates 
in this case  a mixture of various  bilocal operators  
in which  $G_{\mu\alpha}( uz)  G_{\nu \beta }( 0)$ is projected onto 
 the structures built from the metric tensor $g$ and the vectors $p$ and  $z$. 

The   results for arbitrary indices $\sigma \rho \mu\lambda$ are  given  below.
We  present them   in the operator form,
however,  the operators that  have the form of a full derivative are abandoned. 
In other words, we keep   only those operators that survive 
in the forward matrix element.  

 The full result for the box correction to the  forward matrix element 
 of the $  G_{\sigma\rho} \widetilde G_{\mu \lambda}$  operator  may be represented by a sum of three terms.
 The first one has 
 $\Gamma(d/2)$ as an overall factor.
\begin{align}
& G_{\sigma\rho}(z)  \widetilde G_{\mu \lambda} (0) \stackrel{\rm Box, 1}{\longrightarrow}   \frac{g^2N_c \Gamma(d/2)}{4\pi^2\left(z_3^2\right)^{d/2}} 
 \left(\epsilon_{\sigma\rho\mu z}z_\lambda  -\epsilon_{\sigma\rho\lambda z}z_\mu       \right) \int_0^1 \dd u  \frac{\bar u^3}{3}  G_{z\xi}(uz)  G_{z}^{\ \xi}(0)  + \ldots \ .
\end{align}
On the right-hand side here and in the next two equations 
we omit  terms containing  an extra ${\cal O} (z^2)$ factor,
operators with  $D_\nu G^{\mu \nu}$ or with   more than two gluon fields.

The second term is proportional to 
$\Gamma(d/2-1)$\\
\begin{align}
& G_{\sigma\rho}(z)  \widetilde G_{\mu \lambda} (0) \stackrel{\rm Box, 2}{\longrightarrow}    \frac{g^2N_c \Gamma(d/2-1)}{8\pi^2\left(z_3^2\right)^{d/2-1}}   \int_0^1 \dd u  \left\{ \vphantom{\frac{1}{2}}  \epsilon_{\sigma\rho\mu\lambda} \frac{\bar u^3}{3} G_{z\xi}(uz)  G_{z }^{\ \xi}(0) 
 \right. \nn &\quad \left.
 -\frac{\bar u^3}{3} \epsilon_{\sigma\rho\lambda}^{\ \ \ \ \nu} G_{z\nu}( uz) G_{z\mu}(0) 
  \right.
 \left. - \left (2u\bar u+\frac{\bar u^3}{3} \right )
   \epsilon_{\sigma\rho\lambda}^{\ \ \ \ \nu}G_{z\mu}(uz)  G_{z\nu}( 0) 
 \right. \nn 
 &\quad 
 + \bar u^2  \left( \vphantom{\frac{1}{2}} \right. \epsilon_{\sigma\rho z}^{\ \ \ \ \eta}   G_{\lambda\eta}(uz)G_{z\mu}(0)  
  -\epsilon_{\sigma\rho}^{\ \ \ \nu\eta}z_\mu  G_{z\nu}(uz) G_{\lambda\eta}(0) 
   \left. \vphantom{\frac{1}{2}} \right)    \nn 
  &\quad +\bar u(1+u) \left(\vphantom{\frac{1}{2}} \right.
   \epsilon_{\sigma\rho z}^{\ \ \ \ \eta}   G_{z\mu}(uz) G_{\lambda\eta}(0) 
  -  \epsilon_{\sigma\rho}^{\ \ \ \nu\eta}  z_\mu G_{\lambda\eta}(uz)G_{z\nu}(0)  
    \left. \vphantom{\frac{1}{2}} \right)      \nn 
 &\quad +  \left (\frac{\bar u^2}{2} -\frac{\bar u^3}{3} \right ) 
  \left( \vphantom{\frac{1}{2} }\right. \epsilon_{\sigma\rho z \lambda}  \Big ( G_{\mu\xi}(uz)  G_{z }^{\ \xi}(0) 
   + 
   G_{z\xi}(uz)  G_{\mu }^{\ \xi}(0) \Big ) 
        \nn & \quad \hspace{25mm} 
  -\epsilon_{\sigma\rho\lambda}^{\ \ \ \ \nu}z_\mu \Big (  G_{\nu\xi}(uz)  G_{z }^{\ \xi}(0) 
  + 
  G_{z\xi}(uz)  G_{\nu }^{\ \xi}(0)   
  \Big )
   \left. \vphantom{\frac{1}{2} } 
      \right)  \nn 
&\quad  + 2\bar u 
\epsilon_{\sigma\rho z}^{\ \ \ \ \eta} z_\mu   G_{\lambda\xi}(uz) G_{\eta }^{\ \xi}(0)
 \left. -\frac{\bar u^3}{6}  
 \epsilon_{\sigma\rho z \lambda} z_\mu   
 G_{\eta\xi}(uz) G^{\eta\xi}(0) \right\} \     - \{\lambda \leftrightarrow \mu \} \ 
  + \ldots \ .
\end{align}
The third term  is proportional to 
$\Gamma(d/2-2)$:
\begin{align}
&G_{\sigma\rho}(z)  \widetilde G_{\mu \lambda} (0) \stackrel{\rm Box, 3}{\longrightarrow}  \frac{1}{2}\epsilon_{\sigma\rho}^{\ \ \nu\eta}\frac{g^2N_c \Gamma(d/2-2)}{8\pi^2\left(z_3^2\right)^{d/2-2}} \int_0^1 \dd u  
 \left\{\vphantom{\frac{1}{2}} \right.   
 -2 \bar u
 G_{\lambda\eta}(uz)G_{\mu\nu}(0)
 \nn & 
  - u    G_{\mu\lambda}(uz)G_{\nu\eta}(0)   
+  \bar u(1/2-u)    G_{\nu\eta}(uz)  G_{\mu\lambda}(0)+\bar u(1/2+u)    G_{\mu\lambda}(uz) G_{\nu\eta}(0)\nn
&  + 
{\bar uu^2}
 g_{\lambda\eta} \left(  G_{\mu\xi}(uz) G_{\nu }^{\ \xi}(0)  
 +
G_{\nu\xi}(uz)  G_{\mu }^{\ \xi}(0) 
 \right)        
+\bar u \left( g_{\mu\nu}  G_{\lambda\xi}(uz) G_{\eta }^{\ \xi}(0) 
 - g_{\mu\eta}  G_{\lambda\xi}(uz) G_{\nu }^{\ \xi}(0) 
  \right)  
\nn &\hspace{39mm}   - \frac{\bar u^3}{6}
 g_{\mu\nu} g_{\lambda\eta}  
  G_{\zeta\xi}(uz) G^{\zeta\xi}(0)   \left. \vphantom{\frac{1}{2}} \right\} 
 - \{\lambda \leftrightarrow \mu \} \  + \ldots \ . 
\end{align}
We use here the notation 
$ \epsilon_{z\alpha \beta \gamma} =z^\mu  \epsilon_{\mu \alpha \beta \gamma}$, etc.

In practice, however, one may  only need the projections of these expressions onto 
 particular combinations of indices corresponding to matrix elements 
$ \widetilde{ M}_{0 i;   0 i}  + \widetilde{ M}_{ij;   ij}$, \mbox{$\widetilde{ M}_{3 i;  3 i  } - \widetilde{ M}_{ij;   ij} $,} $ \widetilde{ M}_{0 i;  3 i  } $ 
 and  $  \widetilde{ M}_{3 i;  0 i  } $
 that  contain the ``twist-2''  invariant amplitude $\widetilde{\mathcal{M}}_{ps}^{(+)}-
 \nu  \widetilde{\mathcal{M}}_{pp}$  and are   listed in Eqs.  (\ref{00m}),  (\ref{3i3iA}),   (\ref{0i3iA}) and (\ref{3i0iA}).

\section{Matching relations}

As discussed already, 
the sum 
  $  \widetilde{ M}_{00} \equiv  \widetilde{ M}_{0 i;   0 i}  + \widetilde{ M}_{ij;   ij}  $
  contains only the invariant amplitudes $ \widetilde{\mathcal{M}}_{sp}^{(+)}$ and $ \widetilde{\mathcal{M}}_{pp} $ entering in the 
  ``twist-2'' combination \mbox{$\widetilde{\mathcal{M}}_{sp}^{(+)}  - \nu    \widetilde{\mathcal{M}}_{pp} $} .
  Moreover,  since
\begin{align}
 \widetilde{ M}_{00}  =  - 2  p_3 p_0 \left  [ \widetilde{\mathcal{M}}_{sp}^{(+)} -\nu    \widetilde{\mathcal{M}}_{pp}   -\nu    \widetilde{\mathcal{M}}_{pp}m^2/p_3^2  \right ]  \  , 
 \label{M004}
 \end{align}
  the ratio $ \widetilde{ M}_{00}/(-2p_3p_0)$ tends to  $ \widetilde{\mathcal{M}}_{sp}^{(+)} -\nu    \widetilde{\mathcal{M}}_{pp} $ 
 for large $p_3$ at fixed $\nu$. Other combinations of matrix elements, namely, (\ref{3i3iA}),   (\ref{0i3iA}) and (\ref{3i0iA}),
 contain extra  ``contaminating'' invariant amplitudes, like $\widetilde{\mathcal{M}}_{sz} ^{(+)} $, $\widetilde{\mathcal{M}}_{pz} ^{(+)} $,
$ \widetilde{\mathcal{M}}_{zz}$, etc. 
 For this reason,  the combination $ \widetilde{ M}_{0 i;   0 i}  + \widetilde{ M}_{ij;   ij}  $
 is the primary object of the ongoing lattice studies of the polarized gluon distribution. 
 \newpage 
 
 \subsection{Total one-loop correction}
 
  Combining all the  one-loop corrections for the relevant operator (assuming that it is inserted into a forward matrix element $\langle \ldots \rangle$)  we get 
 
 \begin{align}
\langle G_{0 i} & ( z)    \widetilde G_{ 0i}(0) +  G_{ij}( z)   \widetilde G_{ ij}(0)  \rangle \nn
\to &  \frac{g^2N_c }{8\pi^2} \left[{4\over 3}\left( \frac{1}{ \epsilon _{\text{UV}}}+\log\left(z_3^2  \mu^2 \frac{e^{2\gamma_E}}{4}\right) \right)+2 \right]
\left \langle  G_{0i}( z) \widetilde G_{0i}(0) +  G_{ij}( z) \widetilde G_{ij}(0) \right \rangle \nn
& + \frac{g^2N_c }{8\pi^2} \int_0^1 \dd u   \left( {1\over \bar u} - \bar u \right)_+
 \left  \langle G_{0i}( uz) \widetilde G_{0i}(0) +  G_{ij}( uz) \widetilde G_{ij}(0) \right \rangle  \nn
& +  \frac{g^2N_c }{8\pi^2}   \int_0^1 \dd u  \left\{ \vphantom{\frac{1}{2}}  \right. \bar u^2
 \left  \langle G_{  0i}(uz)\widetilde G_{0i }(0) +G_{ i j}(uz)\widetilde G_{ij }(0) \right \rangle 
\nn  
& -\bar u(1+u) 
\left  \langle G_{  3i}(uz)\widetilde G_{3i }(0)  +2G_{ 3  0 }(uz) \widetilde  G_{ 30}(0) \right \rangle
\left. \vphantom{\frac{1}{2}} \right\}  \nn
& +\frac{g^2N_c }{8\pi^2}  \int_0^1 \dd u  \left(\left( \frac{1}{\epsilon_{IR}} -\log\left(z_3^2 \mu^2 \frac{e^{2\gamma_E}}{4}\right)\right)\left(2\bar u u + 2\left[\frac{u}{\bar u}-u\right]_+ + {1\over2}\left({\beta_0\over N_c}-6 \right) \delta(\bar u)\right)
 \right.  \nn &\left. 
- \left[\frac{4 u}{\bar u}
+\frac{4\log (1-u)}{\bar u}\right]_+\right) 
  \left  \langle G_{0 i}(u z)   \widetilde G_{ 0i}(0) +  G_{ij}(u z)   \widetilde G_{ ij}(0)\right \rangle  \nn
& +\frac{g^2N_c }{8\pi^2}\left( \frac{1}{\epsilon_{IR}} -\log\left(z_3^2 \mu^2 \frac{e^{2\gamma_E}}{4}\right)\right) \int_0^1 \dd u \,  2\bar u u 
\left  \langle G_{  3i}(uz)\widetilde G_{3i }(0) +2G_{ 3  0 }(uz) \widetilde  G_{ 30}(0)\right \rangle
\end{align}


Using  the relations in Eqs. (\ref{0=3}) and (\ref{ij=03}) we  change $\langle G_{  3i}(uz)\widetilde G_{3i }(0) +2G_{ 3  0 }(uz) \widetilde  G_{ 30}(0) \rangle $
into $\langle G_{0 i}(u z)   \widetilde G_{ 0i}(0) +  G_{ij}(u z)   \widetilde G_{ ij}(0)\rangle $
and  write everything in terms of the latter.
Switching 
to matrix elements, we get 
 \begin{align}
&{\widetilde M}_{0i;0i}(z,p) + {\widetilde M}_{ij;ij}(z,p) \nn
&\to  \frac{g^2N_c }{8\pi^2} \left[{4\over 3}\left( \frac{1}{ \epsilon _{\text{UV}}}+\log\left(z_3^2 \mu^2 \frac{e^{2\gamma_E}}{4}\right) \right) + 2 \right] \left(  {\widetilde M}_{0i;0i}(z,p) + {\widetilde M}_{ij;ij}(z,p) \right)\nn
&\quad +  \frac{g^2N_c }{8\pi^2}   \int_0^1 \dd u  \left\{-2\bar uu + \left( {1\over \bar u} - \bar u \right)_+-4 \left[\frac{u+\log (1-u)}{\bar u}\right]_+ \right.\nn
&\quad \left. + \left( \frac{1}{\epsilon_{IR}} -\log\left(z_3^2 \mu^2 \frac{e^{2\gamma_E}}{4}\right)\right)\left[\left\{4u\bar u +2\left[u^2/\bar u \right]_+   \right\} +{1\over2}\left({\beta_0\over N_c}-6 \right) \delta(\bar u)\right]  \right\} \nn
&\quad \times \left(  {\widetilde M}_{0i;0i}(uz,p) + {\widetilde M}_{ij;ij}(uz,p)\right) \ .
\end{align}


\subsection{Gluon-quark mixing}

  \begin{figure}[h]
   \centerline{\includegraphics[width=1.7in]{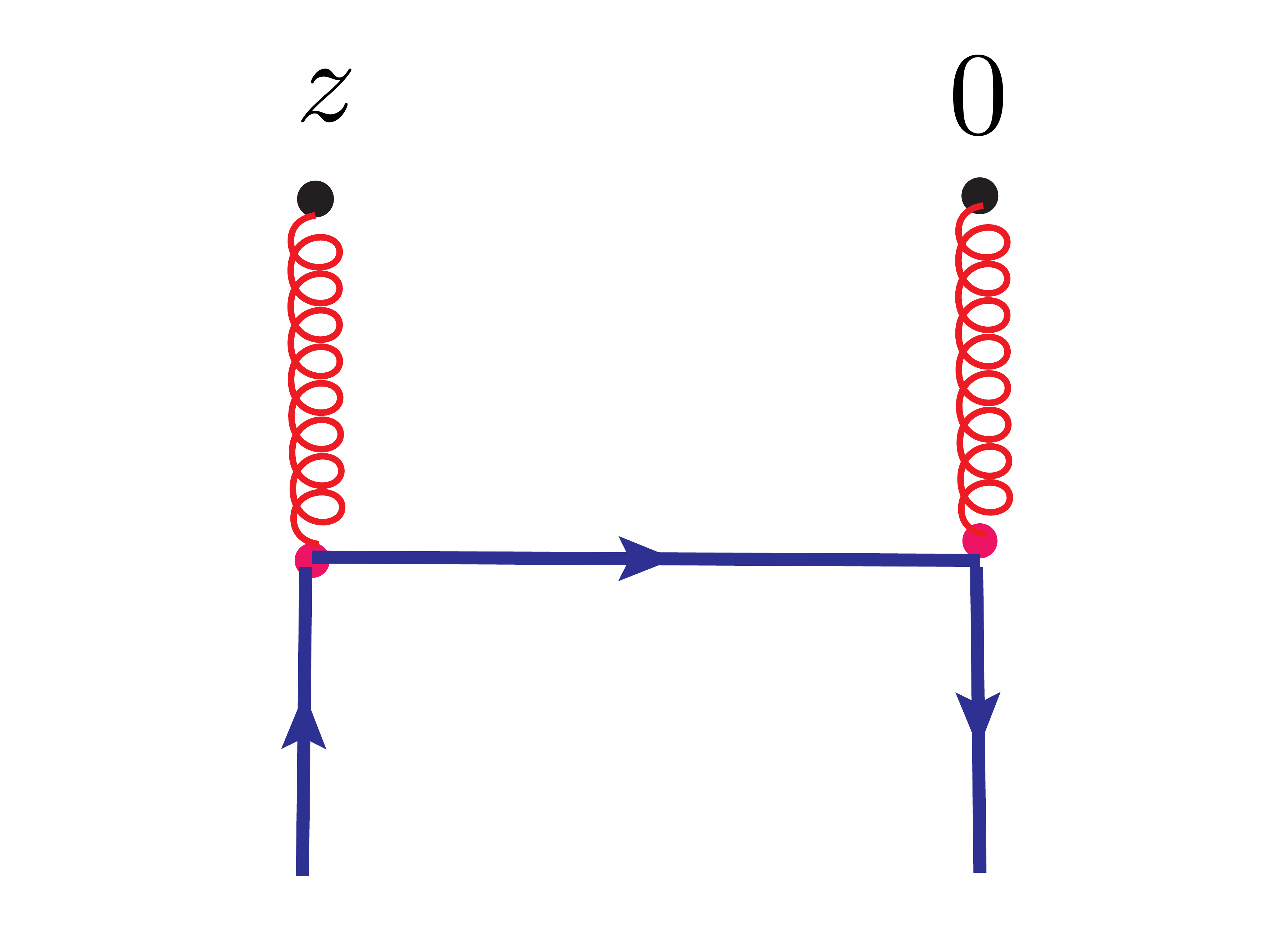}}
        \vspace{-3mm}
   \caption  {Gluon-quark mixing diagram.     \label{gluqum}}
   \end{figure}
   
In addition to the gluon-gluon transitions, we also need to include the contribution from gluon-quark mixing. The result that correspons to $ {\widetilde M}_{0i;0i} + {\widetilde M}_{ij;ij} $ in the $\overline{\rm MS}$ scheme at the operator level is: 
\begin{align}
  - \frac{ g^2 C_F}{8\pi^2}  \int_0^1 \dd u    ~2\bar uu  \,  \partial_0 \mathcal{O}_q^0 \left( uz \right)  -   \frac{ g^2 C_F}{8\pi^2}   \log \left(z_3^2 \mu^2 { e^{2 \gamma_E} \over 4 } \right) \int_0^1 \dd u  \left( 1-\bar u^2 \right) \partial_0 \mathcal{O}_q^0 \left( uz \right).
\label{gqOp}
\end{align}
The singlet combination of quark fields is defined as
\begin{align}
 \mathcal{O}_q^0 \left( z \right) &= {1 \over 2} \sum_f \left( \bar\psi_f( z)   \gamma^0 \gamma_5\psi_f(0)  +\bar\psi_f( 0)     \gamma^0 \gamma_5\psi_f(z) \right) ,
\end{align}
with $f$ numerating quark flavors. Since $\mathcal{O}_q^0$ is even in $z$, the matrix element can be parametrized by
\begin{align}
\bra{p,s} \mathcal{O}_q^0 \left( z \right) \ket{p,s} &= - \, 2 i p_3 \int_{0}^1 \dd x\cos\left( xpz\right) \Delta f_S\left(x\right).
\end{align}
Then, applying the time derivative, we have:
\begin{align}
\partial_0\bra{p,s} \mathcal{O}_q^0 \left( z \right) \ket{p,s} = -\, 2p_0p_3 \, i  \Delta \mathcal{I}_S \left( \nu \right)   ,
\end{align}
where $\nu = - (zp)$, as usual, and
\begin{align}
\Delta \mathcal{I}_S \left( \nu \right) &= \int_{0}^1 x \sin\left( x\nu\right) \Delta f_S\left(x\right).
\end{align}
Applying this parametrization to Eq. (\ref{gqOp}), we obtain:
\begin{align}
 \bra{p,s} & G_{0i}\left( z \right) \widetilde G_{0i} \left( 0 \right) \ket{p,s} + \bra{p,s}G_{ij}\left( z \right) \widetilde G_{ij} \left( 0 \right) \ket{p,s} \nn 
 \to &   2 p_0p_3  \frac{ g^2 C_F}{8\pi^2}  \int_0^1 \dd u  \left[ \log \left(z_3^2 \mu^2 { e^{2 \gamma_E} \over 4 } \right)  \widetilde {\cal B}_{gq} (u) + 2\bar uu    \right] i \Delta \mathcal{I}_S \left(u \nu \right) ,
\label{00gq}
\end{align}
with the $gq$ component of the evolution kernel given by $\widetilde {\cal B}_{gq} (u) =  1- (1-u)^2$.

\subsection{Building reduced Ioffe-time pseudodistribution}
 
 A disadvantage of $ \widetilde{ M}_{00} (z_3,p_3)$ is that it is proportional to $p_3$ for small momenta $p_3$,
 and one cannot use $\widetilde{ M}_{00} (z_3,p_3=0) $
 in the denominator of the ratio defining the 
 reduced pseudo-ITD, like it is done in Eq. (\ref{redm0}).
 To overcome this difficulty, we propose to form the ratio of $ \widetilde{ M}_{00} (z_3,p_3)$ and the 
 $p_3=0$ value of the 
 unpolarized  matrix element
$M_{00}  \equiv  { M}_{0 i;  i 0 }  +  { M}_{ i j; j i} $
of the operator  \mbox{${ G}_{0 i} G_{  i 0 }  +  { G}_{ i j} G_{ j i}$ }
discussed in Ref.\cite{Balitsky:2019krf}.
 As established there, at  the tree level, $M_{00}(z_3,p_3) =   2   p_0^2  \mathcal{M}_{pp}(\nu, z_3^2)$, 
 with  the invariant amplitude $ { \mathcal{M}}_{pp}(\nu, z_3^2)$ 
being  proportional 
to the  pseudo-ITD for the  unpolarized gluon density  $xf_g(x)$ divided  by $\langle x_g \rangle$. 
 Thus,  we are going to  consider the pseudo-ITD $\widetilde{\mathfrak{M} }\left( \nu, z_3^2 \right)$ defined by 
 \begin{align}
\widetilde{\mathfrak{M}} \left( \nu, z_3^2 \right) \equiv  i  \frac{ \{\widetilde{ M}_{00} \left( z_3, p_3 \right)/p_3p_0\} /Z_{\rm L} (z_3/a_L)}{ \{{M}_{00} \left( z_3,p_3=0 \right)/m^2\} } ~.
\end{align}
The factor $i$ is included in  view of Eq. (\ref{Ipnu}), and the factor 
$1/Z_{\rm L} (z_3/a_L)$  (defined by Eq. (\ref{vAD}))  is introduced to cancel the UV  logarithmic vertex AD 
of the $\widetilde{ M}_{00} $ matrix element. 

As we discussed, the main reason for taking the ratio is to cancel 
the factor $Z_{\rm lin} (z_3^2/a^2)$ generated by  linear divergence in the gluon-link self-energy.
This factor is the same in $\widetilde{ M}_{00} \left( z_3, p_3 \right)$ and in ${M}_{00} \left( z_3,p_3=0 \right )$,
so this factor cancels in the ratio.
Furthermore, the denominator factor does not have 
DGLAP evolution logarithms, hence the DGLAP structure 
of  $\widetilde{\mathfrak{M}} \left( \nu, z_3^2 \right)$ is determined 
by DGLAP logarithms of the numerator factor $\widetilde{ M}_{00} \left( z_3, p_3 \right)$.

  Using the   results  of our  calculations for the one-loop corrections to the combinations  
 ${\widetilde M}_{0i;0i}(z,p) + {\widetilde M}_{ij;ij}(z,p) $ and ${ M}_{0 i;  i0 } (z_3, p_3=0) +  { M}_{ i j;  ji} (z_3, p_3=0)$,  and 
neglecting the additional term in Eq. (\ref{00+ii}) with factor $z_3^2/\nu$,
  we obtain 
 the matching  relation 
 \begin{align}
\widetilde{\mathfrak{M}} \left( \nu, z_3^2 \right) \langle x_g \rangle_{\mu^2} &=   
 {{\cal I}_p (\nu, \mu^2 )   } -  \frac{\alpha_s N_c }{2\pi}   \int_0^1 \dd u\,  {{\cal I}_p (u\nu, \mu^2 ) 
 }  \left\{ \log\left(z_3^2 \mu^2 \frac{e^{2\gamma_E}}{4}\right)
\right. \nn & \left. 
\left ( \left [{2u^2 \over \bar u} + 4u\bar u  \right]_+ - \left(\frac 12  + \frac 43  {\langle x_S \rangle_{\mu^2}
 \over  \langle x_g \rangle_{\mu^2} } \right) \delta( \bar u ) \right) \right.\nn
&\quad \left.  +4 \left[\frac{u+\log (1-u)}{\bar u}\right]_+ - \left( {1\over \bar u} - \bar u \right)_+  -\frac 12 \delta(\bar u) +2\bar uu \right\}      \nn
& \quad - \frac{ \alpha_s C_F}{2\pi}  \int_0^1 \dd u \, { \Delta \mathcal{I}_S \left(u \nu,\mu^2 \right)  } 
 \left\{\log \left(z_3^2 \mu^2 { e^{2 \gamma_E} \over 4 } \right)  \widetilde {\cal B}_{gq} (u) + 2\bar uu    \right\} 
 \label{matching}
\end{align}
between   the ``lattice function''   
 $ \widetilde{\mathfrak  M}(  \nu,z_3^2) $ and the polarized light-cone ITDs  for gluons $  {\cal I}_p (\nu,\mu^2)$
 and for quarks $ \Delta \mathcal{I}_S \left(\nu,\mu^2 \right)$.  The factor 
  \begin{align}
  \langle x_g \rangle_{\mu^2} \equiv \int_0^1  \dd x \, xf_g(x,\mu^2) 
  \end{align}
    has the meaning of the  fraction of the hadron 
 momentum carried by the gluons, while  
   \begin{align}
  \langle x_S \rangle_{\mu^2} \equiv \sum_f \int_0^1  \dd x \, x\left (f_f (x,\mu^2) +f_{\bar f} (x,\mu^2)\right )
  \end{align}  
 corresponds to the  fraction of the hadron 
 momentum carried by the singlet quarks.  Note that $ \left[{ 2u^2 / \bar u }  +4u\bar u \right]_+ $ coincides for $u\neq 1$ with 
 the   $gg$-part of the Altarelli-Parisi kernel for polarized gluon distribution $x \Delta g(x,\mu^2 )$ 
 (see, e.g.,  Ref. \cite{Balitsky:1997mj}).

 
  Eq. (\ref{matching})  allows one to extract just the shape of  the polarized gluon distribution.
 Its normalization, i.e., the magnitude  of $ \langle x_g \rangle_{\mu^2} $  must  be 
 taken from  an independent  lattice calculation, similar to that performed 
 in Ref. \cite{Yang:2018bft}.
 The singlet quark function $\Delta {\cal I}_S (w \nu, \mu^2)$ that appears in the  
 ${\cal O} (\alpha_s)$ correction and $ \langle x_S \rangle_{\mu^2} $ should be also calculated (or estimated)     
 independently.

 Using Eq. (\ref{glPDFsin}) allows us to write (\ref{matching}) directly in terms of the LC polarized gluon distribution:
  \begin{align}
  ~\widetilde{\mathfrak{M}} \left( \nu, z_3^2 \right) = & \int_0^1 \dd x { x \Delta g(x,\mu^2 )\over \expval{x_g}_{\mu^2}} \widetilde R_{gg} \left(x\nu, z_3^2\mu^2 \right) 
  \nn &
 + \int_0^1 \dd x { x \Delta f_S(x,\mu^2 )\over \expval{x_g}_{\mu^2}} \widetilde R_{gq} \left(x\nu, z_3^2\mu^2 \right) ,
 \end{align}
 where the gluon-gluon kernel $ \widetilde R_{gg} $ is given by 
 \begin{align}
 \widetilde R_{gg}  \left(x\nu, z_3^2\mu^2 \right) 
 =  \sin(x\nu) 
  - & \frac{g^2N_c }{8\pi^2}\int_0^1   \dd u \sin(ux\nu)  \left\{ \log\left(z_3^2 \mu^2 \frac{e^{2\gamma_E}}{4}\right)
\right. \nn & \left. 
\left ( \left [{2u^2 \over \bar u} + 4u\bar u  \right]_+ - \left(\frac 12  + \frac 43  {\langle x_S \rangle_{\mu^2}
 \over  \langle x_g \rangle_{\mu^2} } \right) \delta( \bar u ) \right) \right.\nn
&\quad \left.  +4 \left[\frac{u+\log (1-u)}{\bar u}\right]_+ - \left( {1\over \bar u} - \bar u \right)_+  -\frac 12 \delta(\bar u) +2\bar uu \right\}   ,
 \end{align}
 and  the gluon-quark  kernel $ \widetilde R_{gq} $ is 
 \begin{align}
  \widetilde R_{gq} & \left(x\nu, z_3^2\mu^2 \right) \nn
  = &  -  \frac{ \alpha_s C_F}{2\pi}  \int_0^1 \dd u \sin(ux\nu)\left\{\log \left(z_3^2 \mu^2 { e^{2 \gamma_E} \over 4 } \right)  \widetilde {\cal B}_{gq} (u) + 2\bar uu    \right\} .
 \end{align}

 \section{ Summary}

In  this paper, we  formulated the basic points 
of the pseudo-PDF approach to  lattice  calculation of polarized  gluon PDFs.  
In particular, we have presented the   results  of our calculations of the one-loop 
corrections for the bilocal  $G_{\mu \alpha}(z)  \widetilde G_{\lambda \beta}(0)$ correlator of gluonic fields. 
We gave  the expressions  for  a general  situation when all four indices are arbitrary,
and also specified them for  combinations of indices giving    three matrix elements
that contain the structures corresponding to  twist-2 invariant amplitude related to the polarized PDF. 
We   have  studied the evolution properties  of  these matrix elements,
and derived matching relations between Euclidean and light-cone Ioffe-time distributions
that are necessary for extraction of the polarized gluon distributions from the lattice data.

{\bf Acknowledgements.} We  thank K. Orginos,  \mbox{J.-W. Qiu}, D. Richards, R. Sufian and T. Khan    for 
interest   in  our   work and  discussions. This work is supported by Jefferson Science Associates,
 LLC under  U.S. DOE Contract \#DE-AC05-06OR23177
 and by U.S. DOE Grant \#DE-FG02-97ER41028. 
 
 \newpage 

 \bibliography{Polar.bib}
\bibliographystyle{jhep}

    \end{document}